\documentclass[english,11pt]{revtex4-1}
\usepackage{graphicx}
\usepackage{hyperref}
\usepackage{cancel}
\usepackage{amssymb}
\usepackage{textcomp}
\usepackage{amsmath}
\usepackage{bm}
\usepackage{times}
\usepackage{epsfig}
\usepackage{color}
\usepackage{mathrsfs}
\usepackage{esint}

\makeatletter
\@ifundefined{textcolor}{}
{%
 \definecolor{BLACK}{gray}{0}
 \definecolor{WHITE}{gray}{1}
 \definecolor{RED}{rgb}{1,0,0}
 \definecolor{GREEN}{rgb}{0,1,0}
 \definecolor{BLUE}{rgb}{0,0,1}
 \definecolor{CYAN}{cmyk}{1,0,0,0}
 \definecolor{MAGENTA}{cmyk}{0,1,0,0}
 \definecolor{YELLOW}{cmyk}{0,0,1,0}
 }

\makeatletter

\newcommand{\be}{\begin{equation}}
\newcommand{\ee}{\end{equation}}
\newcommand{\ba}{\begin{align}}
\newcommand{\ea}{\end{align}}
\def\bea{\begin{eqnarray}}
\def\eea{\end{eqnarray}}

\usepackage{slashed}

\makeatother

\usepackage{babel}

\newcommand{\eV}{{\rm eV}}
\newcommand{\MeV}{{\rm MeV}}
\newcommand{\GeV}{{\rm GeV}}
\newcommand{\TeV}{{\rm TeV}}
\newcommand{\fb}{{\rm fb}}

\begin{document}

\title{The $B-L$ Scotogenic Models for Dirac Neutrino Masses}

\author{Weijian Wang}
\email{wjnwang96@aliyun.com}
\affiliation{Department of Physics, North China Electric Power
University, Baoding 071003,China}
\author{Ruihong Wang}
\email{rhongwang@163.com}
\affiliation{Department of Physics, North China Electric Power
University, Baoding 071003,China}
\affiliation{College of Information Science $\&$ Technology, Hebei Agricultural University, Baoding,China}
\author{Zhi-Long Han}
\email{sps\_hanzl@ujn.edu.cn}
\affiliation{School of Physics and Technology, University of Jinan, Jinan, Shandong 250022, China}
\author{Jin-Zhong Han}
\affiliation{School of Physics and Telecommunications Engineering,
Zhoukou Normal University, Henan, 466001, China}
\email{jinzhonghan@aliyun.com}

\begin{abstract}
 We construct the one-loop and two-loop scotogenic models for Dirac
neutrino mass generation in the context of $U(1)_{B-L}$ extensions
of standard model. It is indicated that the total number of
intermediate fermion singlets is uniquely fixed by anomaly free
condition and the new particles may have exotic $B-L$ charges so
that the direct SM Yukawa mass term $\bar{\nu}_L\nu_R\overline{\phi^0}$
 and the Majorana mass term $(m_N/2)\overline{\nu_R^C}\nu_R$ are
naturally forbidden. After the spontaneous breaking of $U(1)_{B-L}$ symmetry,
the discrete $Z_{2}$ or $Z_{3}$ symmetry appears as the residual
symmetry and give rise to the stability of intermediated fields as
DM candidate. Phenomenological aspects of lepton flavor violation,
DM, leptogenesis and LHC signatures are discussed.
\end{abstract}

\maketitle

\section{Introduction}
The standard model(SM) needs extensions to incorporate two important
missing pieces: the tiny neutrino masses and the cosmological  dark
matter (DM) candidates. The scotogenic model, proposed by
Ma\cite{Ma:2006km}, has recently became an attractive and economical
scenario to accommodate the above two issues in a unified framework.
The main idea is based on the assumption that the DM candidates can
serve as intermediate messengers propagating inside the loop diagram
in neutrino mass generation. Classical examples are the Ma's
one-loop model\cite{Ma:2006km} and two-loop model\cite{Ma:2007gq}.
Some representative variations are found in
Refs.~\cite{Ma:2008cu,Ma:2013yga,Law:2013saa,3loop,Aoki:2008av,
3loop2,Ahriche:2014oda,Baek:2015mna,Ma:2015xla,Fraser:2015mhb,
Ding:2016wbd,Ahriche:2016cio,Ahriche:2016ixu,Cheung:2016frv,
Nomura:2016dnf,Lu:2016ucn,Guo:2016dzl,Lu:2016dbc,Liu:2016mpf,
Gu:2016xno,Ko:2017quv,Cheung:2017efc,Lee:2017ekw,Baek:2017qos,
CarcamoHernandez:2016pdu,Sierra:2016rcz,Simoes:2017kqb,Nomura:2017ezy,
Nomura:2017emk}.
In these models, the stability of DM is usually guaranteed by imposing
the odd parity under $\emph{ad hoc}$ $Z_{2}$ or $Z_{3}$ symmetry.
The origin of discrete symmetry is still unknown. An attractive
scenario, known as Krauss-Wilczek mechanism~\cite{wel}, is that the
discrete symmetry appears as the residual symmetry which originates
from the spontaneous symmetry breaking(SSB) of a continuous gauge
symmetry at high scale. The simplest and well-studied gauge extension
of SM is that of $U(1)_{B-L}$, which was first realized within the
framework of left-right symmetric models~\cite{Pati:1974yy,Mohapatra:1974gc,
Senjanovic:1975rk,Mohapatra:1980qe}. Following this spirit, several loop-induced
Majorana neutrino mass models were constructed based on gauged
$U(1)_{B-L}$ symmetry~\cite{Kanemura:2011vm,Chang:2011kv,Kanemura:2011mw,Kanemura:2014rpa,
Wang:2015saa,Ma:2016nnn,Ho:2016aye,Seto:2016pks}.
In these works, exotic $B-L$ charges are assigned to new particles
to satisfy the anomalies cancelation condition. By taking
appropriate charge assignment, the residual discrete $Z_{2}(Z_3)$
symmetry arises after the SSB of $U(1)_{B-L}$ symmetry. Then the
lightest particles with odd $Z_{2}(Z_3)$ parity can not decay into SM
ingredients, becoming a DM candidate.

On the other hand, the evidences establishing whether neutrinos are
Majorana or Dirac fermion is still missing. If neutrinos are Dirac
fermions, certain new physics beyond the SM should exist to account
for the tiny neutrino mass. Several scotogenic models for Dirac
neutrino masses were
proposed in Refs.~\cite{Gu:2007ug,Farzan:2012sa,Okada:2014vla,
Kanemura:2016ixx,Bonilla:2016diq,Borah:2016zbd,Borah:2017leo}.
The generic one-loop topographies are discussed in Ref.~\cite{Ma:2016mwh} and
subsequently,  specific realizations with $SU(2)_{L}$ multiplets
fields are presented in Ref.~\cite{Wang:2016lve}.  In these models,  two
$ad$ $hoc$ discrete symmetries were introduced, one is responsible
for the absence of SM Yukawa couplings $\bar{\nu}_L\nu_R\overline{\phi^0}$
and the other for the stability of intermediate fields as dark matter(DM).The symmetries could be
discrete $Z_2$\cite{Farzan:2012sa,Ma:2016mwh,Wang:2016lve}, $Z_3$\cite{Ma:2015mjd,Bonilla:2016diq},
or $Z_4$\cite{Heeck:2013rpa,Heeck:2013vha}.

It is natural to ask if the $B-L$ symmetry also shed light on Dirac neutrino mass generation and DM phenomena. Recently several efforts were made at tree level\cite{Ma:2015mjd,Ma:2014qra,Ma:2015raa,Chulia:2016ngi}, and a specific one-loop realization was also proposed based on left-right symmetry scheme\cite{Borah:2017leo}. In this brief article, we propose the $U(1)_{B-L}$ extensions of
scotogenic Dirac neutrino mass models with intermediate Dirac
fermion singlets.  We will systematically discuss the
one- and two-loop realizations for Dirac neutrino masses
with typical topographies respectively.  In these models, a singlet scalar $\sigma$ is responsible for the SSB of gauged $U(1)_{B-L}$ symmetry as well as masses of the heavy intermediate Dirac fermions. To get the Dirac type neutrino mass term, we introduce three right-handed components $\nu_{R}$ and assume that they share the same $B-L$ charges.  The intermediate Dirac fermions are
SM singlets but carry $B-L$ quantum numbers. This implies that the
anomaly cancelations of $[SU(3)_{c}]^{2}\times U(1)_{B-L}$,
$[SU(2)_{L}]^{2}\times U(1)_{B-L}$ and $U(1)_{Y}\times[U(1)_{B-L}]^{2}$ are
automatically satisfied. Thus we only need to consider the
$[U(1)_{B-L}]\times[Gravity]^{2}$ and $[U(1)_{B-L}]^{3}$ anomaly
conditions. Then the effective Dirac neutrino mass term $m_D \bar{\nu}_L \nu_R$ is
induced by SSB of $U(1)_{B-L}$. As we shall see, the discrete
$Z_{2}$ or $Z_{3}$ symmetry could appear as a remnant symmetry of gauged
$U(1)_{B-L}$ symmetry, naturally leading to DM candidates.

In Sec.II, we construct the one/two-loop diagrams for Dirac neutrino mass generation and discuss their validity under $B-L$ anomaly free condition. We consider the phenomenology of the models in Sec.III. A summary is given in Sec. IV.

\section{Model Building}
\subsection{One-loop Scotogenic Model}
Consider first the one-loop scotogenic realization of Dirac neutrino masses.  In the $B-L$ extended scotogenic models, the particle content under $SU(2)_{L}\times U(1)_{Y}\times U(1)_{B-L }$ symmetry is listed as follow
\begin{eqnarray}
&L \sim (2,-1/2, -1),\quad \nu_{R1,2,3}\sim (1,0,Q_{\nu_{R}}),\quad F_{L/Ri}\sim (1,0,Q_{F_{L/R}})
\\\nonumber
&\Phi\sim (2,1/2,0),\quad \eta\sim (2,1/2,Q_{\eta}),\quad \chi\sim(1,0,Q_{\chi}),\quad \sigma\sim (1,0,Q_{\sigma})
\end{eqnarray}
where several Dirac
fermion singlets are added with their chiral components denoted as
$F_{Ri}$ and $F_{Li}$($i=1\cdots n$) respectively. In the scalar sector,
we further add one doublet scalar $\eta$ and one singlet scalar $\chi$.

\begin{figure}[!htbp]
\begin{center}
\includegraphics[width=0.29\linewidth]{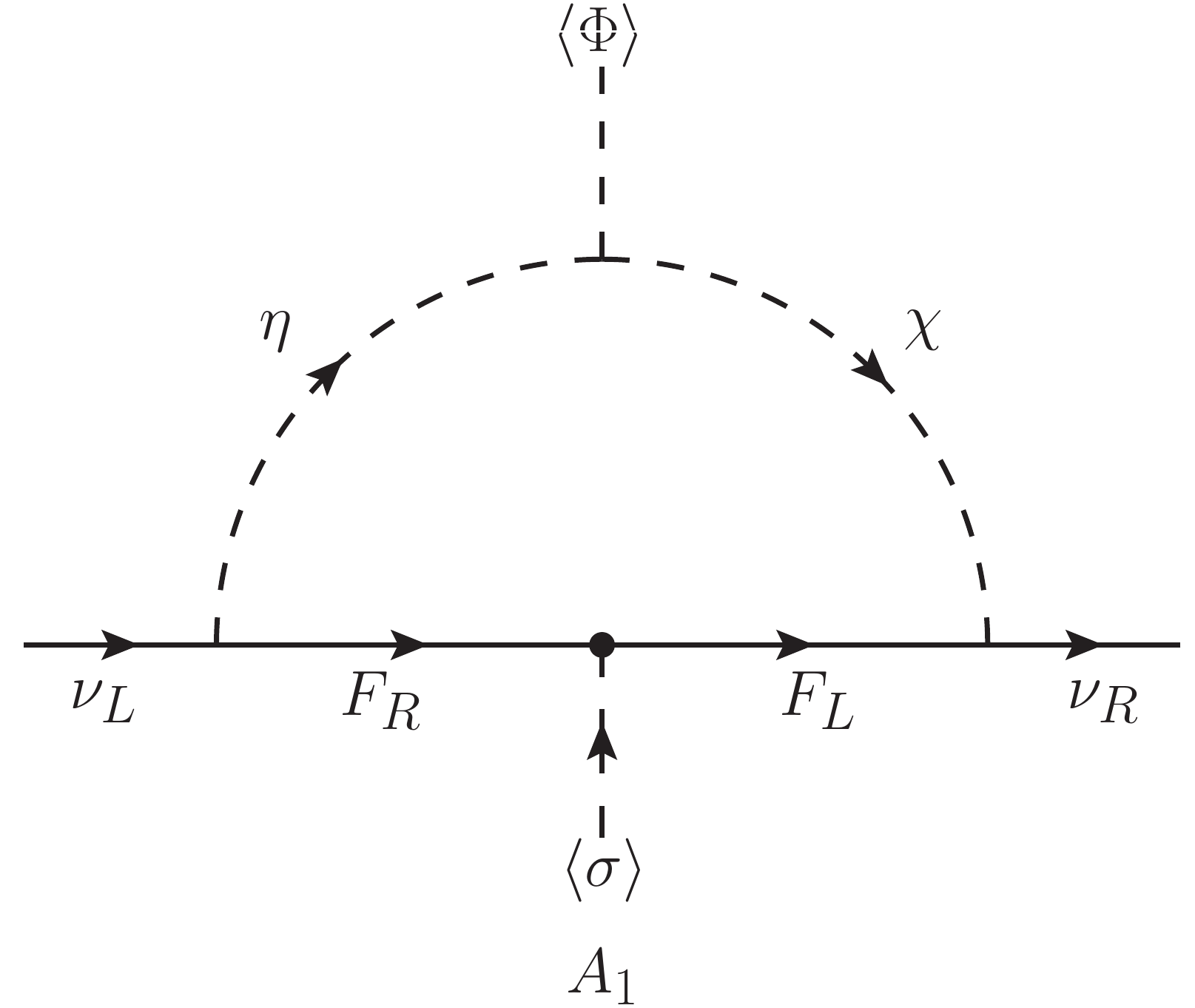}
\includegraphics[width=0.29\linewidth]{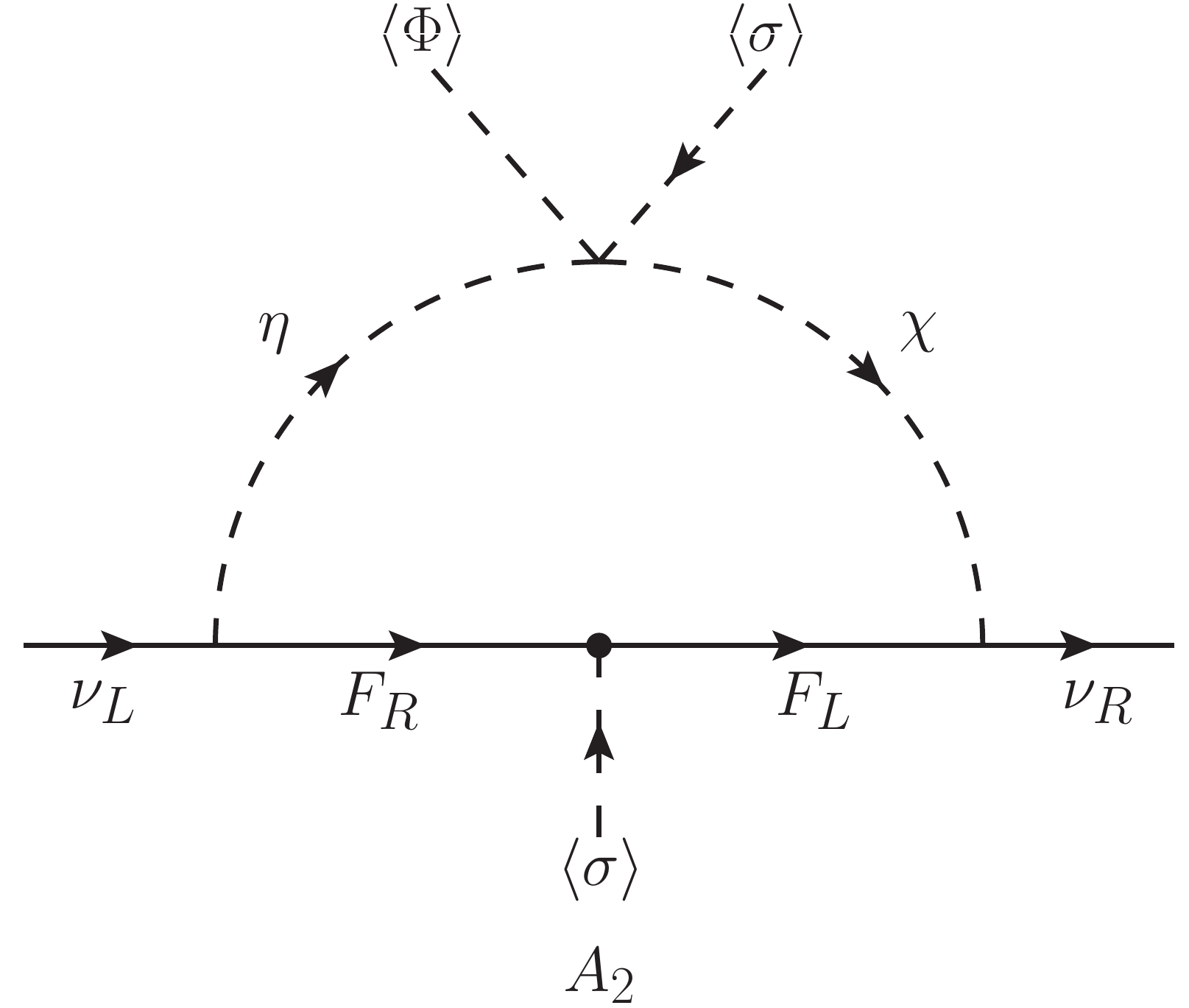}
\includegraphics[width=0.29\linewidth]{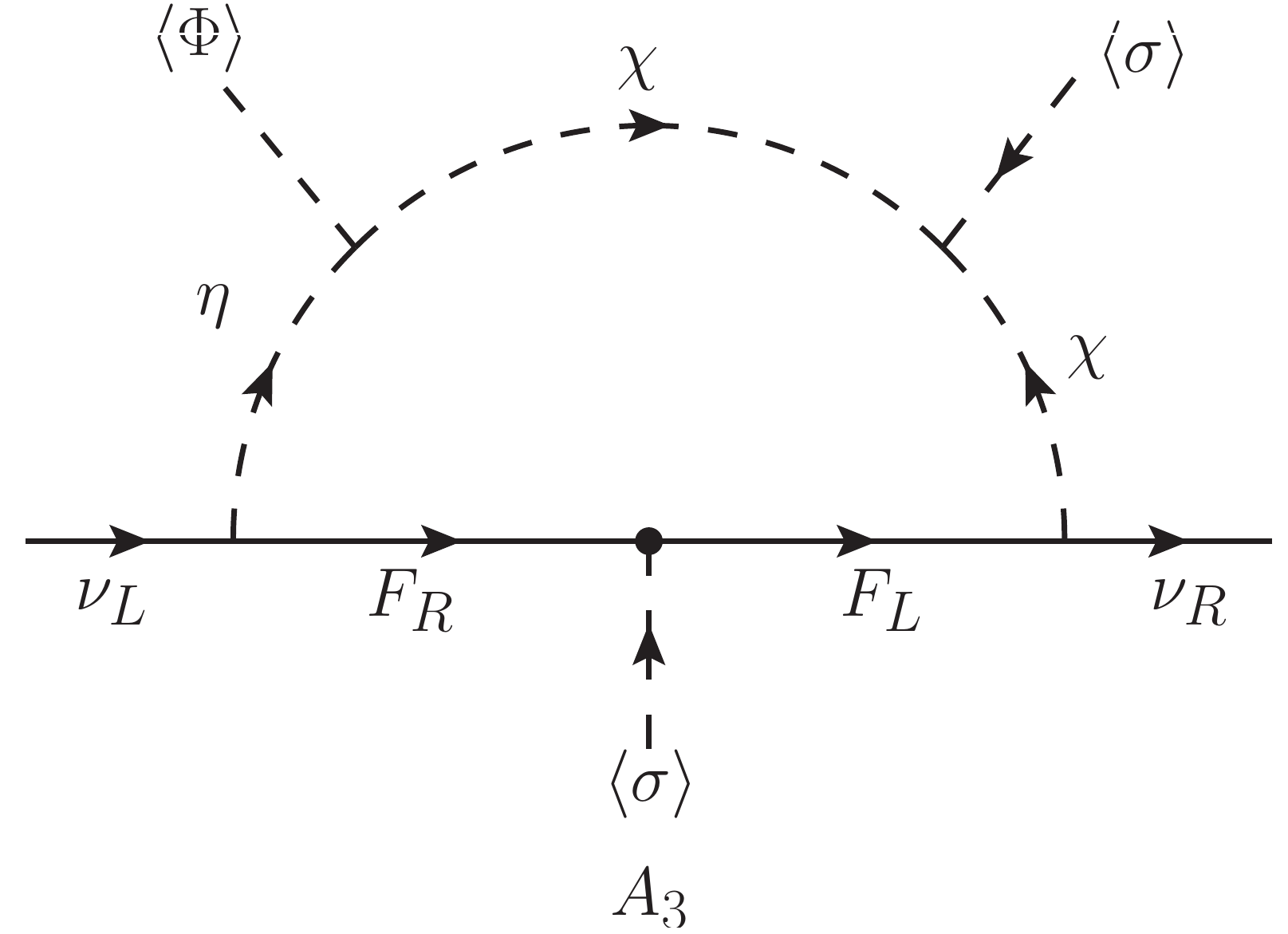}
\includegraphics[width=0.29\linewidth]{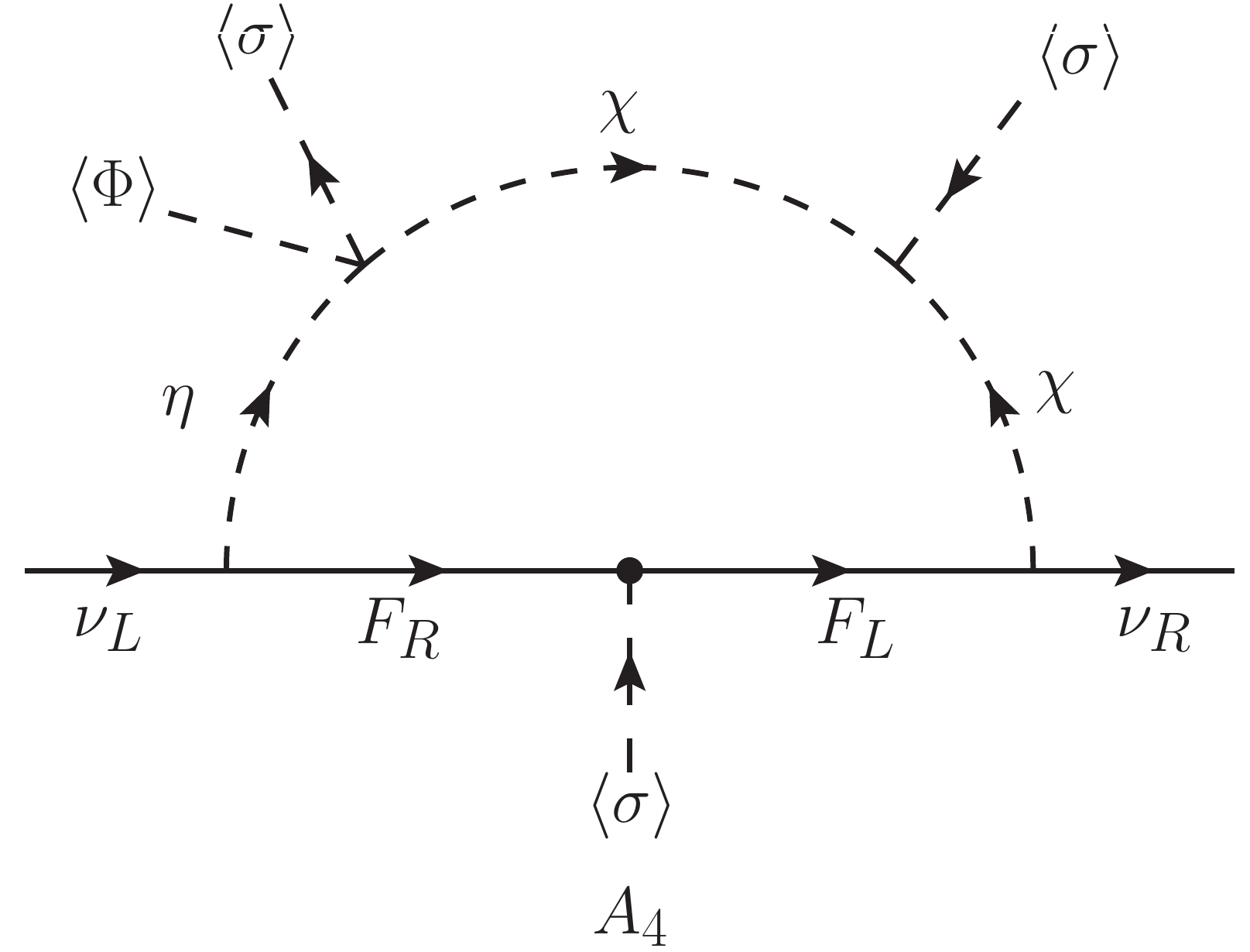}
\includegraphics[width=0.29\linewidth]{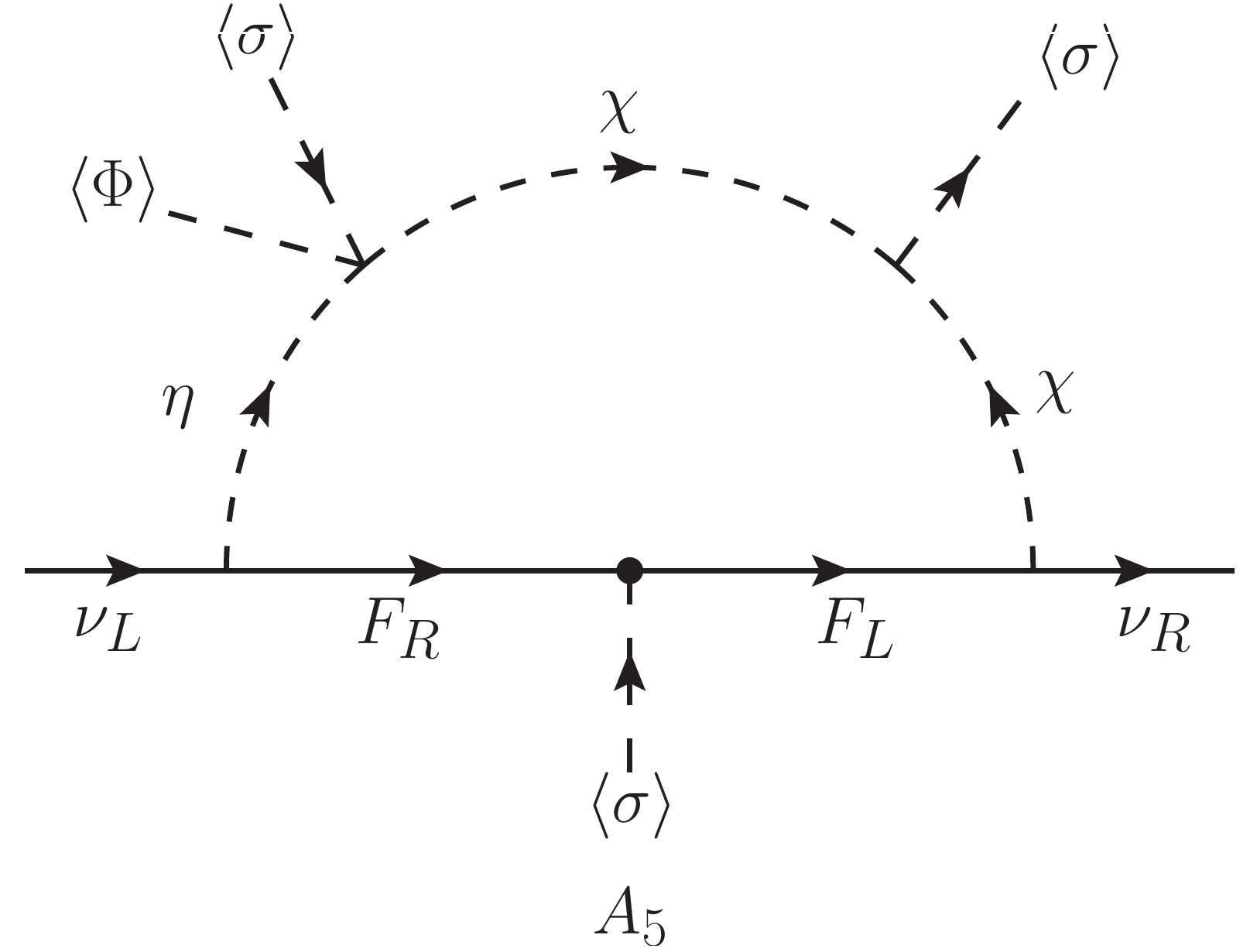}
\includegraphics[width=0.29\linewidth]{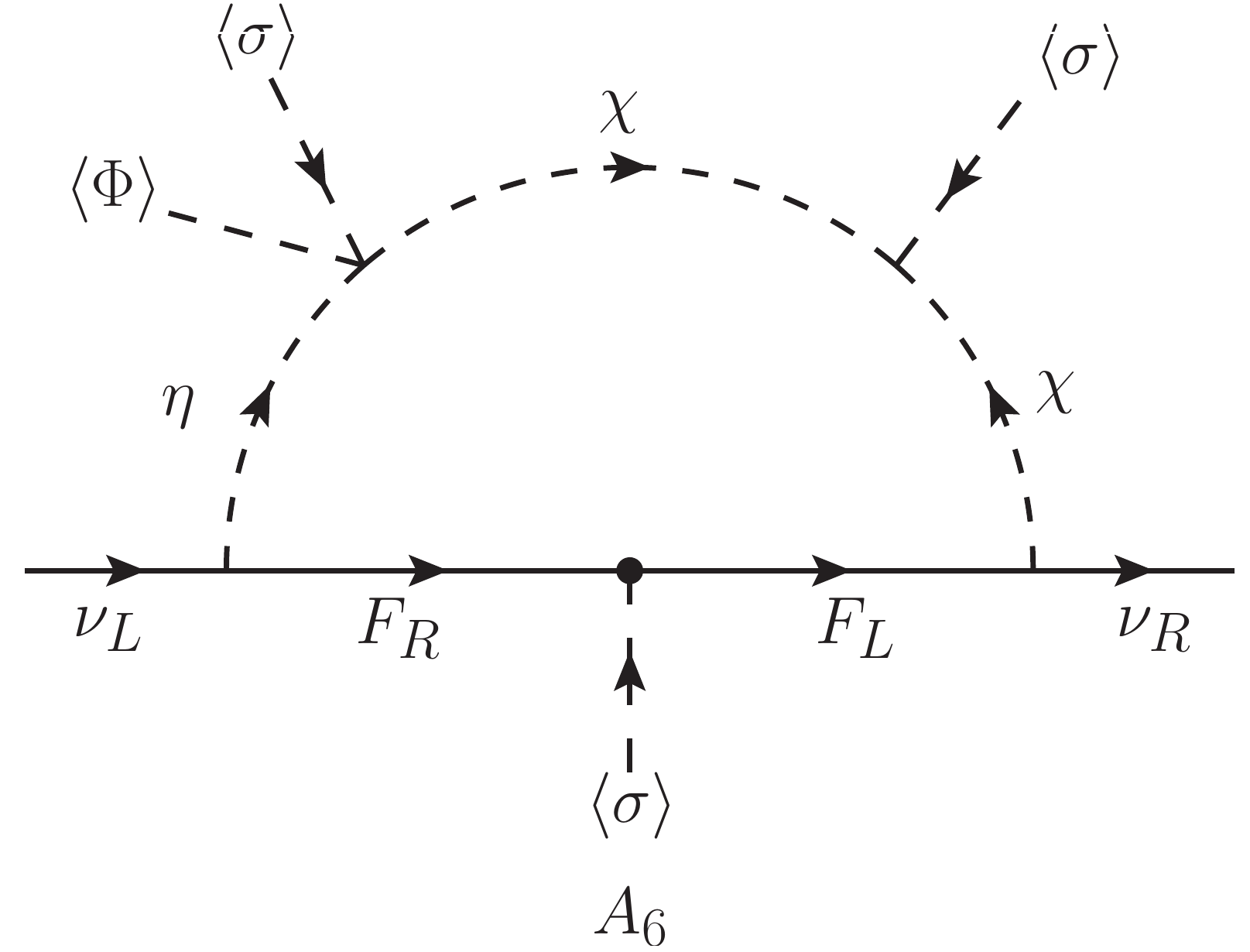}
\includegraphics[width=0.29\linewidth]{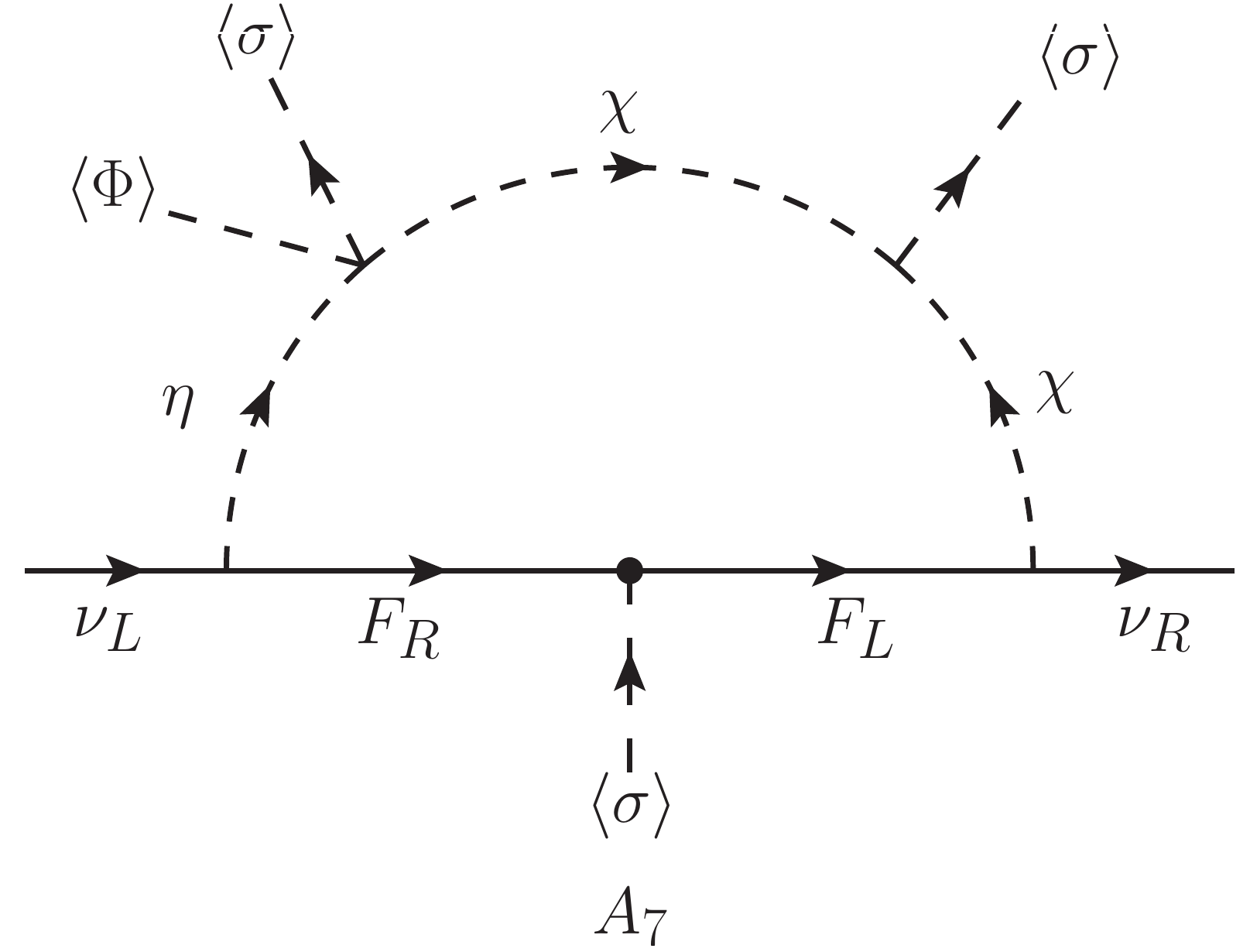}
\end{center}
\caption{ Possible one-loop topological diagrams that can generate
the prototype model given in Ref.~\cite{Farzan:2012sa} after the SSB
of $U(1)_{B-L}$ symmetry \label{A}}
\end{figure}

In the original $Z_{2}$ model \cite{Gu:2007ug,Farzan:2012sa}, $Z_{2}$ odd parity is
assigned to $\nu_{R}$ and intermediated particle fields running in the loop. As a warm up, we start from the simplest $U(1)_{B-L}$ extension. We denote it as $A_{1}$ model with the corresponding Feynman diagrams illustrated as the first diagram in Fig.~\ref{A}. The relevant interactions for radiative Dirac
 neutrino mass generation are given as
\begin{equation}\label{oneflow}
\mathcal{L} ~\supset~ y_{1}\overline{L}F_{R}i\tau_{2}\eta^{\ast}+
y_{2}\overline{\nu_{R}}F_{L}\chi+ f\overline{F_{L}}F_{R}\sigma+\mu
(\Phi^{\dagger}\eta)\chi^{\ast}+\text{h.c.},
\end{equation}
where $L$ is the SM lepton doublet and we omit the summation
indices. In terms of gauged $U(1)_{B-L}$ symmetry, one should consider the
$[U(1)_{B-L}]\times[Gravity]^{2}$ and $[U(1)_{B-L}]^{3}$ anomaly
free conditions
\begin{eqnarray}
&-3-3Q_{\nu_{R}}-nQ_{F_{R}}+nQ_{F_{L}}=0,\\\nonumber
&-3-3Q_{\nu_{R}}^3-nQ_{F_{R}}^3+nQ_{F_{L}}^3=0,
\label{g1}\end{eqnarray}
which, using relevant interactions given in Eq.\eqref{oneflow} , can be solved exactly as
\begin{equation}
n=3, \quad\quad Q_{F_{R}}=-Q_{\nu_{R}},\quad\quad Q_{F_{L}}=1
\end{equation}
Given the interations in Eq.\eqref{oneflow}, the charge assignments for other particles are listed in the $A_{1}$ row in Table. I. Therefore the total number of heavy fermions is fixed by the anomaly
free conditions and the $B-L$ charge assignments for all new particles
 are determined in terms of  free parameter $Q_{\nu_{R}}$. Let us now discuss precisely what values $Q_{\nu_R}$ can be taken. First,  the condition $Q_{\nu_{R}}\neq-1$ should also be imposed to forbid the SM direct Yukawa coupling term $\bar{\nu}_L\nu_R\overline{\phi^0}$.
Second, forbidding Majorana mass terms
$(m_{R})\overline{\nu_{R}^C}\nu_{R}$, $\sigma\overline{\nu_{R}^C}\nu_{R}$ and $\sigma^{\ast}\overline{\nu_{R}^C}\nu_{R}$ requires $Q_{\nu_{R}}\neq0, -1/3$ and $ 1$ respectively (note that $Q_{\sigma}=Q_{\nu_{R}}+1$ for $A_{1}$ model).  Third,  to generate a purely loop-induced neutrino mass term, $Q_{\sigma}$ and $Q_{\chi}$($=Q_{\nu_{R}}-1$) appropriately assigned so that
$\sigma^{k}\chi$ and $(\sigma^{\ast})^{k}\chi$($k=1,2,3$) terms,
which cause the VEV of $\chi$, are forbidden. This further requires
$Q_{\nu_{R}}\neq 0, -1/3,-1/2,-2$ and $-3$.  Similarly, the $(\Phi^{\dag}\eta)\sigma^{k}$ and $(\Phi^{\dag}\eta)(\sigma^{\ast})^{k} (k=1,2)$ should also be avoid to generate the VEV of $\eta$, leading to $Q_{\nu_{R}}\neq 0, -1/3,-3$. Once an appropriate $Q_{\nu_{R}}$ is taken,
the residual $Z_{2}$ symmetry appears in Eq.\eqref{oneflow}, under
which the parity is odd for inert particles ($\eta,\chi,F_{L/R}$)
and even for all other particles.

We now consider other possible realizations. In the scalar sector,
the interactions relevant to radiative neutrino mass generation are
given by
\begin{eqnarray}\label{mt}
\mathcal{L}_{S} ~\supset~
&(\Phi^{\dag}\eta)\chi,\quad(\Phi^{\dag}\eta)\chi^{\ast},\quad(\Phi^{\dag}\eta)\chi\sigma,
\quad(\Phi^{\dag}\eta)\chi^{\ast}\sigma,\\\nonumber
&(\Phi^{\dag}\eta)\chi\sigma^{\ast},\quad(\Phi^{\dag}\eta)\chi^{\ast}\sigma^{\ast},
\quad\chi^{2}\sigma,\quad\chi^{2}\sigma^{\ast},\quad+ \text{h.c.}
\end{eqnarray}
Taking appropriate charge assignment, at least one $\eta-\chi$ mixing term given in
Eq.~\eqref{mt} should be selected to build the model. All the seven possible
topological diagrams(denoted as $A_{1}-A_{7}$) are depicted in Fig.~\ref{A},
where we have already discussed the specific model $A_{1}$ above.

Under the gauged $U(1)_{B-L}$ symmetry, the quantum numbers of new
particles are required to satisfy the anomaly free conditions.  We
summarize the $B-L$ quantum number assignments for each diagram in
Table~\ref{ca}. We have checked that among the seven models, five of them
($A_{1}$,$A_{2}$, $A_{4}$, $A_{5}$ and $A_{6}$) are suitable for the gauged
$B-L$ extension. For each available model, the total number of intermediate
fermions $F_{R/L}$ is uniquely determined by the anomaly free condition
of $[U(1)_{B-L}]\times[Gravity]^{2}$. The $B-L$ quantum number of
$A_{1}$ and $A_{2}$ model can not be uniquely fixed and we choose
$Q_{\nu_{R}}$ as the variable. If $\chi$ linear terms are forbidden
by appropriate $Q_{\nu_{R}}$ assignment, the residual $Z_{2}$ symmetry arises
after the SSB of $U(1)_{B-L}$. Thus the lightest particle with odd
$Z_{2}$ parity can serve as a DM candidate.

\begin{table}
\begin{tabular}{c c c c c c c c c}
\hline\hline Models & \quad$n$& $\nu_{R}$& $F_{R}$ & $F_{L}$
 &$\eta$ &$\chi$ & $\sigma$ & Scalar interactions \\
\hline
$A_{1}$& \quad 3 &$x$&$-x$& 1& $x-1$& $x-1$ &$x+1$& $(\Phi^{\dag}\eta)\chi^{\ast}$\\
$A_{2}$& \quad 6 &$x$&$\frac{\pm z-x-1}{4}$& $\frac{\pm z+x+1}{4}$&$\frac{\mp z+x-3}{4}$& $\frac{\mp z+3x-1}{4}$ &$\frac{x+1}{2}$&$(\Phi^{\dag}\eta)\chi^{\ast}\sigma$\\
$A_{3}$& \quad $\times$ &$\times$&$\times$& $\times$& $\times$& $\times$ &$\times$&  $(\Phi^{\dag}\eta)\chi^{\ast},\chi^{2}\sigma$\\
$A_{4}$& \quad 3 &$3$&$-3$& $1$& $2$& $-2$ &$4$&  $(\Phi^{\dag}\eta)\chi^{\ast}\sigma^{\ast},\chi^{2}\sigma$ \\
$A_{5}$& \quad 3 &$\frac{1}{3}$&$-\frac{1}{3}$& $1$& $-\frac{2}{3}$& $\frac{2}{3}$ &$\frac{4}{3}$&  $(\Phi^{\dag}\eta)\chi^{\ast}\sigma,\chi^{2}\sigma^{\ast}$\\
$A_{6}$& \quad 9 &$\frac{23}{13}$&$\frac{5}{13}$& $\frac{17}{13}$& $-\frac{18}{13}$& $-\frac{6}{13}$ &$\frac{12}{13}$&  $(\Phi^{\dag}\eta)\chi^{\ast}\sigma,\chi^{2}\sigma$\\
$A_{7}$& \quad $\times$ &$\times$&$\times$& $\times$& $\times$& $\times$ &$\times$&  $(\Phi^{\dag}\eta)\chi^{\ast},\chi^{2}\sigma^{\ast}$\\
\hline\hline
\end{tabular}
\caption{$B-L$ charge assignments for new particles in each one-loop
models. In A2 model, we set $z\equiv (5x^{2}-6x+5)^{1/2}$. The
symbol ``$\times$'' means that no appropriate charge assignment are available to  meet the requirement of
anomaly cancellation}\label{ca}
\end{table}
Compared with $A_{1}$ and $A_{2}$, for models $A_{4}$, $A_{5}$ and
$A_{6}$, the $B-L$ quantum numbers for new particles are fixed
uniquely. This is due to the fact that the interaction
$\chi^{2}\sigma$ ($\chi^{2}\sigma^{\ast}$) contributes an additional
constraint on $Q_{\chi}$ and $Q_{\sigma}$, i.e.,
\begin{equation}
2Q_{\chi}\pm Q_{\sigma}=0.
\end{equation}
The existence of $\chi^{2}\sigma$($\chi^{2}\sigma^{\ast}$) term has
two-fold meanings: (i) that it automatically forbids the $\chi$
linear terms and guarantee the existence of residual $Z_{2}$
symmetry after the SSB of $U(1)_{B-L}$; (ii) that it induces a mass
splitting $\Delta M=\mid M_{\chi_{R}}-M_{\chi_{I}}\mid$ between
the real ($\chi_{R}$) and imaginary part ($\chi_{I}$) of $\chi$.
Provided $\Delta M$ is larger than the DM kinetic energy
$KE_{D}\sim\mathcal{O}$(100) KeV, the tree-level DM-nucleon scattering
via the $U(1)_{B-L}$ gauge boson $Z'$ and SM $Z$ boson exchange
 (due to the mixing between $\eta$ and $\chi$) are
kinematically forbidden, thus a $\chi_{R}/\chi_I$ dominated DM is
expectable through the scalar singlet $\sigma$ or SM Higgs portal.

One recalls that in the prototype scotogenic Dirac model \cite{Farzan:2012sa}
with sizable Yukawa couplings, a relatively small coupling constants
of $\eta-\chi$ mixing terms is required to reproduce the scale of
neutrino masses. To rationalize such a unnaturally small coupling,
an extra soften broken symmetry is added\cite{Farzan:2012sa}. We
emphasize that the fine tuning can be relaxed in $A_{4}-A_{6}$
models with the help of double suppression from $\eta-\chi$ and
$\chi_{R}-\chi_{I}$ mixing interactions. Takeing $A_{5}$ model as an
example, with scalar interactions $\lambda(\Phi^{\dag}\eta)\chi^{\ast}\sigma$
and $\mu_\chi\chi^{2}\sigma^{\ast}$, the radiative neutrino mass is evaluated
as
\begin{equation}
m_{\nu}\simeq\frac{\lambda y_{1}y_{2}f}{16\pi^{2}}\Big(\frac{\langle
\Phi\rangle\langle\sigma\rangle^{3}}{\Lambda^{4}} \Big)\mu_\chi,
\end{equation}
where $\Lambda\sim m_{\eta}, m_{\chi}^{R},m_{\chi}^{I}$ denotes the
scale of new physics, usually taken to be
$\Lambda\sim\langle\sigma\rangle\sim\mathcal{O}(1)$TeV. Then for
$\lambda\sim y_1\sim y_{2}\sim f\sim10^{-2}$ and
$\mu_\chi\sim \mathcal{O}$($10$)GeV, the neutrino mass scale ($0.1$
eV) can be reproduced.

\subsection{Two-loop Scotogenic Models}
Now let us discuss the two-loop scotogenic realizations of Dirac neutrino
masses.  The simple model with $Z_{3}$ discrete symmetry was proposed
recently\cite{Bonilla:2016diq} where two classes of Dirac fermion
singlets are added. Here we denote the corresponding chiral
components as $F_{R,Li}$($i=1,2\cdots n$) and
$S_{R,Lj}$($j=1,2\cdots m$) respectively. In the scalar sector, we add
one scalar doublet $\eta$, two scalar singlets $\chi$ and $\xi$. In
order to accomplish the $U(1)_{B-L}$ extension, a scalar singlet
$\sigma$ is also added to play the role as $B-L$ symmetry breaking. The
particle content and quantum number assignments under $SU(2)_{L}\times U(1)_{Y}\times U(1)_{B-L }$  gauge symmetry are summarized as follow
\begin{eqnarray}
&L \sim (2,-1/2, -1),\quad \nu_{R1,2,3}\sim (1,0,Q_{\nu_{R}})
\\\nonumber
&F_{L/Ri}\sim (1,0,Q_{F_{L/R}}),\quad S_{L/Rj}\sim (1,0,Q_{S_{L/R}})\\\nonumber
&\Phi\sim (2,1/2,0),\quad \eta\sim (2,1/2,Q_{\eta}),\quad \chi\sim(1,0,Q_{\chi}),\quad \sigma\sim (1,0,Q_{\sigma})
\end{eqnarray}

\begin{figure}
\begin{center}
\includegraphics[width=0.30\linewidth]{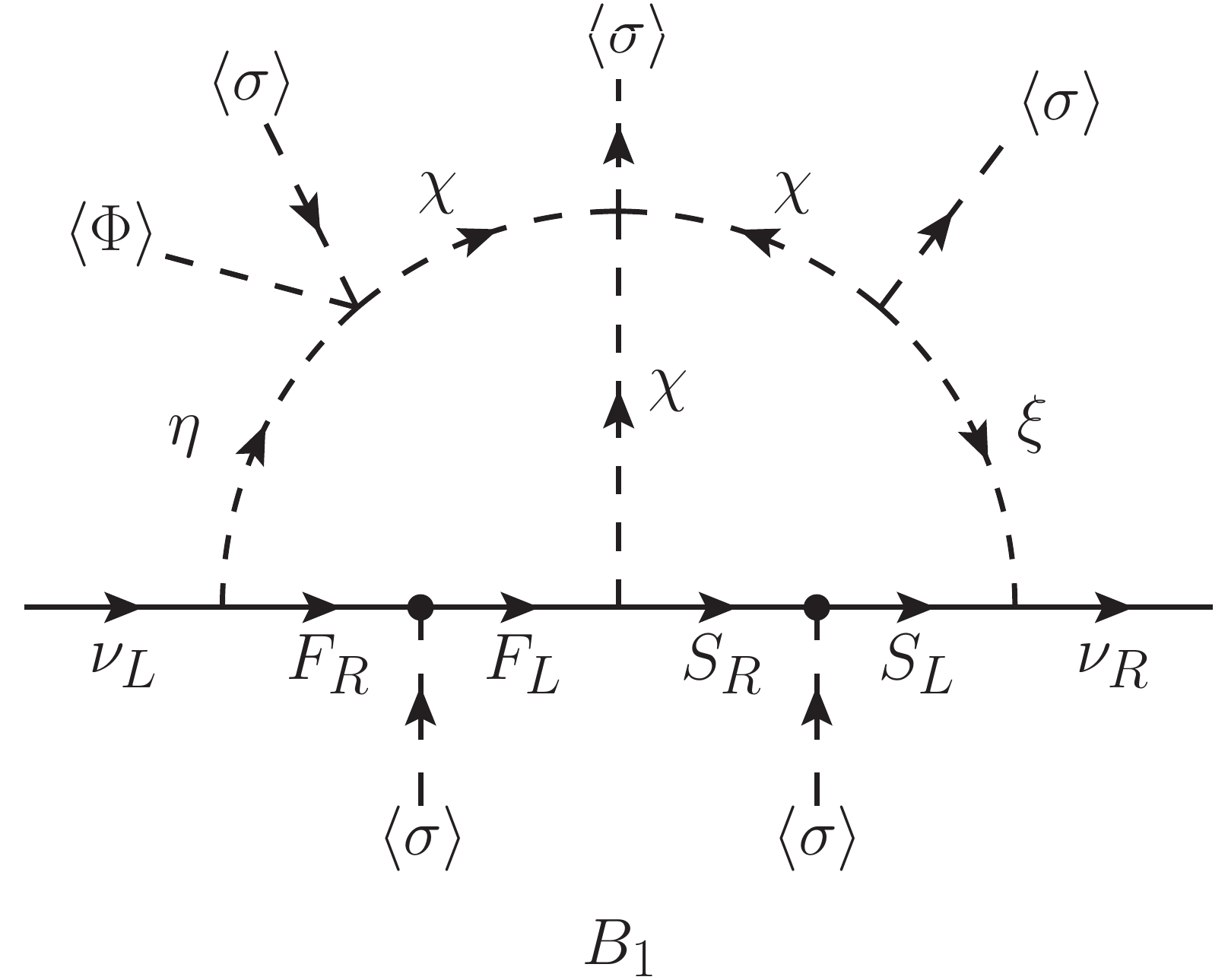}
\includegraphics[width=0.30\linewidth]{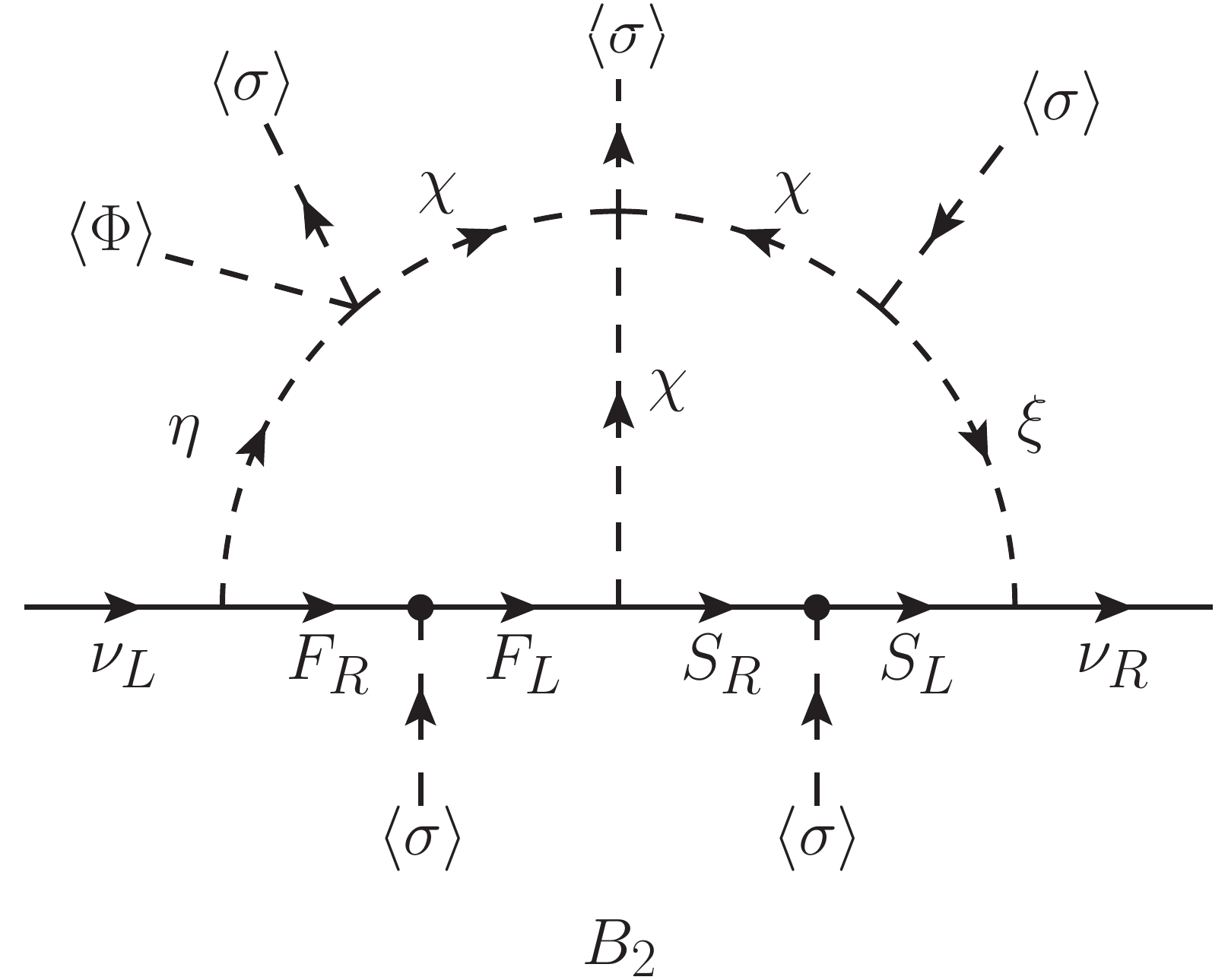}\\
\includegraphics[width=0.30\linewidth]{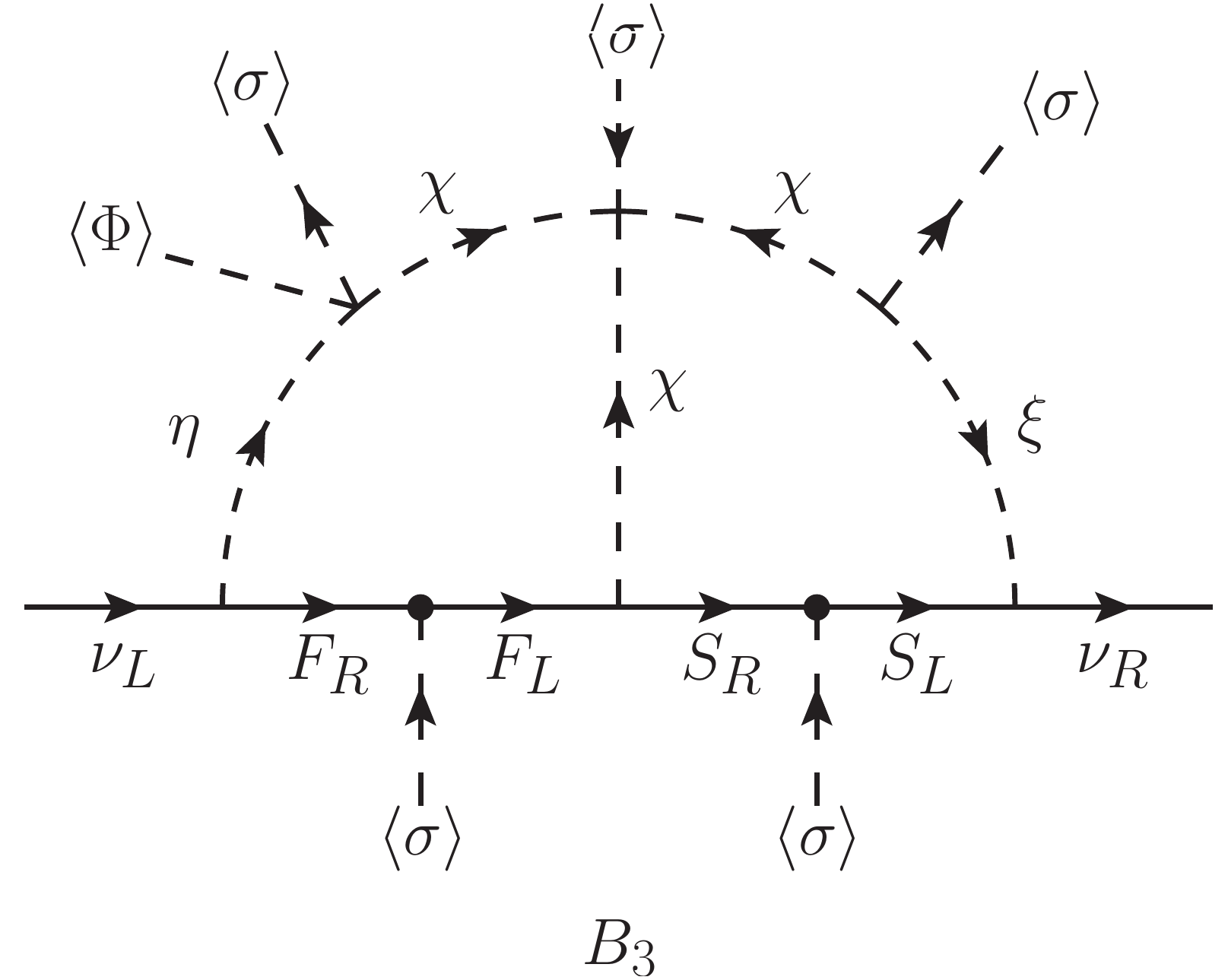}
\includegraphics[width=0.30\linewidth]{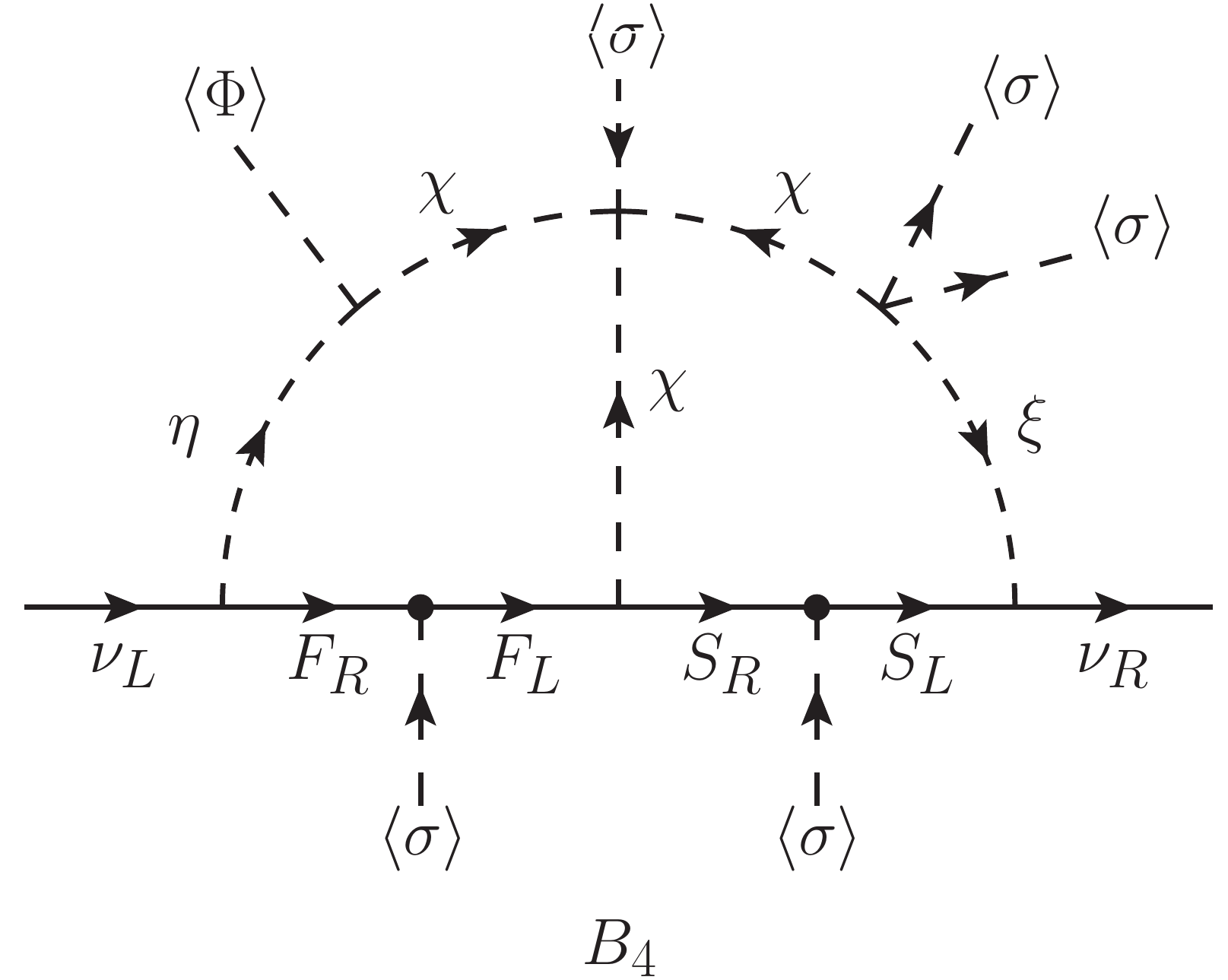}
\end{center}
\caption{ Available two-loop topological diagrams of Dirac neutrino
mass with $n+m=3$.\label{B}}
\end{figure}

Similar as the one-loop cases, the two-loop model can be realized
though various pathways. As an illustration, we start from a simple
$U(1)_{B-L}$ extension (denoted as $B_{1}$) with
topology depicted by the first diagram in Fig.~\ref{B}. The relevant
interactions are
\begin{eqnarray}\label{interaction}
\mathcal{L}~\supset~&y_{1}\overline{L}F_{R}i\tau_{2}\eta^{\ast}+
y_{2}\overline{\nu_{R}}S_{L}\xi+ f_{1}\overline{F_{L}}F_{R}\sigma+
f_{2}\overline{S_{L}}S_{R}\sigma+
h\overline{S}_{R}F_{L}\chi^{\ast}\\\nonumber &+\lambda_{1}
(\Phi^{\dagger}\eta)\chi^{\ast}\sigma+
\lambda_{2}\chi^{3}\sigma^\ast+
\mu_{3}\xi\chi\sigma+\text{h.c.}
\end{eqnarray}
Under gauged $U(1)_{B-L}$ symmetry, the condition of cancelation for
$[U(1)_{B-L}]\times[Gravity]^{2}$ anomaly is given by
\begin{equation}\label{g2}
-3-3Q_{\nu_{R}}-nQ_{F_{R}}+n(Q_{F_{R}}+Q_{\nu_{R}}+1)-mQ_{S_{R}}+m(Q_{S_{R}}+Q_{\nu_{R}}+1)=0
\end{equation}
Notice that $Q_{\nu_{R}}\neq-1$ is
required to forbid $\overline{\nu}_{L}\nu_{R}\overline{\phi^0}$ term.
From Eq.\eqref{g2}, one obtains
\begin{equation}
n+m=3. \label{sb1}
\end{equation}
Clearly, only $(n,m)=(1,2)$
and $(2,1)$ patterns are allowed for model $B_{1}$. In this secnario, the rank of effective neutrino mass matrix is two, implying a vanishing neutrino mass eigenvalue.  Hence the models with condition $n+m=3$ are the minimal two-loop realizations allowed phenomenologically.
The anomaly free
condition of $[U(1)_{B-L}]^{3}$ is given by
\begin{equation}
-3-3Q_{\nu_{R}}^3-nQ_{F_{R}}^3+n(Q_{F_{R}}+Q_{\nu_{R}}+1)^3-mQ_{S_{R}}^{3}+m(Q_{S_{R}}+Q_{\nu_{R}}+1)^{3}=0
\label{b13}
\end{equation}
Taking the interaction terms in
Eq.\eqref{interaction} into account and solving Eq.\eqref{sb1},
\eqref{b13}, we find
\begin{equation}
Q_{\nu_{R}}=\frac{5n-17}{3n+5}.
\end{equation}
Subsequently, the $B-L$ charges of other particles are obtained,
which are shown explicitly in Table~\ref{main1}.

Now we investigate other viable realizations. Without loss of
generality, we focus on the minimal models with three
intermediate fermions, i.e., $n+m=3$.  To generate a residual $Z_3$
discrete symmetry, the $\chi^3\sigma$ or $\chi^3\sigma^\ast$ is needed.
After the SSB of $U(1)_{B-L}$, $\chi$ transforms as $\omega=e^{i 2\pi/3}$
under the residual $Z_3$ symmetry. It is found that only four models are available under the anomaly free condition.
The  corresponding topological diagrams are shown in Fig.~\ref{B}.
Besides $B_{1}$, we denote rest of models as $B_{2}$, $B_{3}$
and $B_{4}$ respectively. Following the same methodology in one-loop case, the $B-L$ charge assignments of new particles for each model
are obtained. The main results are listed in Table~\ref{main1}.
\begin{table}
\begin{tabular}{c c c c c c c c c c c c}
\hline\hline \quad &\quad $(n,m)$ &\quad $\nu_{R1,2,3}$\quad &\quad
$F_{Ri}$ \quad
 &\quad  $F_{Li}$ \quad  & \quad  $S_{Ri}$ \quad  & \quad  $S_{Li}$ \quad  & \quad  $\eta$  \quad & \quad  $\chi$ \quad & \quad  $\xi$  \quad & \quad $\sigma$ & \quad Scalar interactions\\
\hline
$B_{1}$ &$(1,2)$ & $-\frac{1}{11}$& $-\frac{13}{33}$& $\frac{17}{33}$& $\frac{7}{33}$& $\frac{37}{33}$& $-\frac{20}{33}$ & $\frac{10}{33}$ & $-\frac{40}{33}$& \quad $\frac{10}{11}$ &\quad$(\Phi^{\dagger}\eta)\chi^{\ast}\sigma$,$\chi^{3}\sigma^{\ast}$,$\chi\xi\sigma$\\
     &$(2,1)$ & $0$& $-\frac{1}{3}$& $\frac{2}{3}$& $\frac{1}{3}$& $\frac{4}{3}$& $-\frac{2}{3}$ & $\frac{1}{3}$ & $-\frac{4}{3}$& \quad $1$&\quad$(\Phi^{\dagger}\eta)\chi^{\ast}\sigma$,$\chi^{3}\sigma^{\ast}$,$\chi\xi\sigma$ \\
\hline
$B_{2}$ &$(1,2)$ & $\times$& $\times$& $\times$& $\times$& $\times$& $\times$ & $\times$ & $\times$& \quad $\times$&\quad$\times$\\
     &$(2,1)$ & $-11$& $\frac{37}{3}$& $\frac{7}{3}$& $\frac{17}{3}$& $-\frac{13}{3}$ & $-\frac{40}{3}$ & $-\frac{10}{3}$& \quad $-\frac{20}{3}$ & $-10$&\quad$(\Phi^{\dagger}\eta)\chi^{\ast}\sigma^{\ast}$,$\chi^{3}\sigma^{\ast}$,$\chi\xi\sigma^{\ast}$\\
\hline
$B_{3}$ &$(1,2)$ & $-\frac{1}{3}$& $-\frac{13}{9}$& $-\frac{7}{9}$& $-\frac{5}{9}$& $\frac{1}{9}$& $\frac{4}{9}$ & $-\frac{2}{9}$ & $-\frac{4}{9}$& \quad $\frac{2}{3}$&\quad$(\Phi^{\dagger}\eta)\chi^{\ast}\sigma^{\ast}$,$\chi^{3}\sigma$,$\chi\xi\sigma$\\
      &$(2,1)$ & $-3$& $\frac{1}{3}$& $-\frac{5}{3}$& $-\frac{7}{3}$& $-\frac{13}{3}$ & $-\frac{4}{3}$ & $\frac{2}{3}$& \quad $\frac{4}{3}$ & $-2$&\quad$(\Phi^{\dagger}\eta)\chi^{\ast}\sigma^{\ast}$,$\chi^{3}\sigma$,$\chi\xi\sigma$\\
\hline
$B_{4}$ &$(1,2)$ & $-\frac{3}{14}$& $-\frac{31}{42}$& $\frac{1}{21}$& $\frac{13}{42}$& $\frac{23}{21}$ & $-\frac{11}{42}$ & $-\frac{11}{42}$& \quad $-\frac{55}{42}$& \quad$\frac{11}{14}$&\quad$(\Phi^{\dagger}\eta)\chi^{\ast}$,$\chi^{3}\sigma$,$\chi\xi\sigma^{2}$\\
      &$(2,1)$ & $-\frac{1}{8}$& $-\frac{17}{24}$& $\frac{1}{6}$& $\frac{11}{24}$& $\frac{4}{3}$ & $-\frac{7}{24}$& $-\frac{7}{24}$ & $-\frac{35}{24}$&\quad $\frac{7}{8}$&\quad$(\Phi^{\dagger}\eta)\chi^{\ast}$,$\chi^{3}\sigma$,$\chi\xi\sigma^{2}$\\
\hline\hline
\end{tabular}
\caption{$B-L$ quantum number assignments and relevant scalar
interactions for two-loop models with $n+m=3$. }\label{main1}
\end{table}
Obviously, after $B-L$ breaking, the residual $Z_{3}$ symmetry
arises with
\begin{equation}
 F_{L,Ri}\sim\omega,\quad\quad S_{L,Ri}\sim\omega,\quad\quad\eta, \chi \sim
 \omega,\quad\quad\xi \sim \omega^{2}
\end{equation}

\section{Phenomenology: a case study}
In the following, we consider some phenomenological aspects
of the gauged $B-L$ scotogenic Dirac models. From Table. I,
we can see that besides the $B-L$ charge
and some scalar interactions being different, all the
one-loop models have same interactions as in Eq.~\ref{oneflow}.
Therefore, we can concentrate on the simplest one, i.e., model $A_1$.
As for the two-loop models, phenomenon will be similar
provided the additional $\xi$ and $S_{L,R}$ are heavy enough.

In model $A_1$, the $B-L$ charges of all the additional particles
are determined by $B-L$ charge of right-handed neutrino $Q_{\nu_R}$.
To make sure a residual $Z_2$ symmetry after the breaking of $B-L$,
we fix $Q_{\nu_R}=1/6$ in the following discussion.
The complete gauge invariant scalar potential for model $A_1$ is
\begin{eqnarray}
V&=& -\mu_\Phi \Phi^\dag\Phi + \mu_\eta \eta^\dag\eta + \mu_\chi \chi^*\chi - \mu_\sigma \sigma^*\sigma +\lambda_\Phi (\Phi^\dag\Phi)^2+ \lambda_\eta (\eta^\dag \eta)^2 + \lambda_\chi (\chi^*\chi)^2 \\ \nonumber
&& + \lambda_\sigma (\sigma^*\sigma)^2 + \lambda_{\Phi \eta} (\Phi^\dag\Phi) (\eta^\dag \eta) + \lambda_{\Phi \chi} (\Phi^\dag\Phi) (\chi^*\chi) + \lambda_{\Phi \sigma} (\Phi^\dag\Phi) (\sigma^*\sigma)
\\ \nonumber
&& + \lambda_{\eta \chi} (\eta^\dag \eta) (\chi^*\chi) + \lambda_{\eta \sigma} (\eta^\dag \eta) (\sigma^*\sigma) + \lambda_{\chi\sigma} (\chi^*\chi) (\sigma^*\sigma) +
\left[\mu(\Phi^{\dagger}\eta)\chi^{\ast} + \text{h.c.}\right].
\end{eqnarray}
For the $Z_2$ even scalars, $\phi^0_R$ and $\sigma_R$ mix into
physical scalars $h$ and $H$ with mixing angle $\alpha$. Here,
we regard $h$ as the discovered $125~\GeV$ scalar at LHC
\cite{Aad:2012tfa,Chatrchyan:2012xdj,Aad:2015zhl}. In order
to escape various direct and indirect searches for the scalar $H$
\cite{Robens:2015gla},
a small mixing angle $\sin\alpha=0.01$ is assumed in this work.
Meanwhile, for the $Z_2$ odd scalars $\eta^0$ and $\chi$, they will
mix into physical scalars $H_2^0$ and $H_1^0$ with mixing angle
$\beta$. As shown in Refs.~\cite{Cohen:2011ec,Kakizaki:2016dza},
a small mixing angle, e.g., $\sin\beta\lesssim0.01$ is preferred in
case of scalar DM $H_1^0$. In this paper, we take $\sin\beta=10^{-6}$,
mainly aiming to interpret tiny neutrino masses. And we also have
one pair of $Z_2$ odd charged scalar $H_2^\pm(=\eta^\pm)$.

Given the interactions in Eq.~\ref{oneflow},
the one-loop induced neutrino mass for model $A_1$ is
\begin{equation}\label{mv:loop}
m_{\nu}^{ij}= \frac{\sin2\beta}{32\pi^2}\sum_k y_1^{ik} y_2^{jk*}
M_{Fk}\left[\frac{M_{H_2^0}^2}{M_{H_2^0}^2-M_{Fk}^2}\ln\left(\frac{M_{H_2^0}^2}{M_{Fk}^2}\right)
-\frac{M_{H_1^0}^2}{M_{H_1^0}^2-M_{Fk}^2}\ln\left(\frac{M_{H_1^0}^2}{M_{Fk}^2}\right)\right].
\end{equation}
To give some concrete prediction, we present one promising benchmark point (BP) for model A$_1$
\begin{eqnarray}\label{BP}\nonumber
\sin\beta=10^{-6},|y_{1,2}^{i1}|=10^{-6},|y_{1,2}^{i2,i3}|=0.007, \\
M_{H_1^0}=45~\GeV, M_H=100~\GeV, M_{H_2^\pm,H_2^0}=600~\GeV,\\\nonumber
M_{F1}=M_{F2,F3}/2=5~\TeV,M_{Z^\prime}=4~\TeV,g_{BL}=0.1,
\end{eqnarray}
which could realise $m_\nu\sim0.1~\eV$. For simplicity,
we denote $|y_{1,2}^{i2,i3}|=y$ in the following.

\begin{figure}[!htbp]
	\centering
	\includegraphics[width=0.5\linewidth]{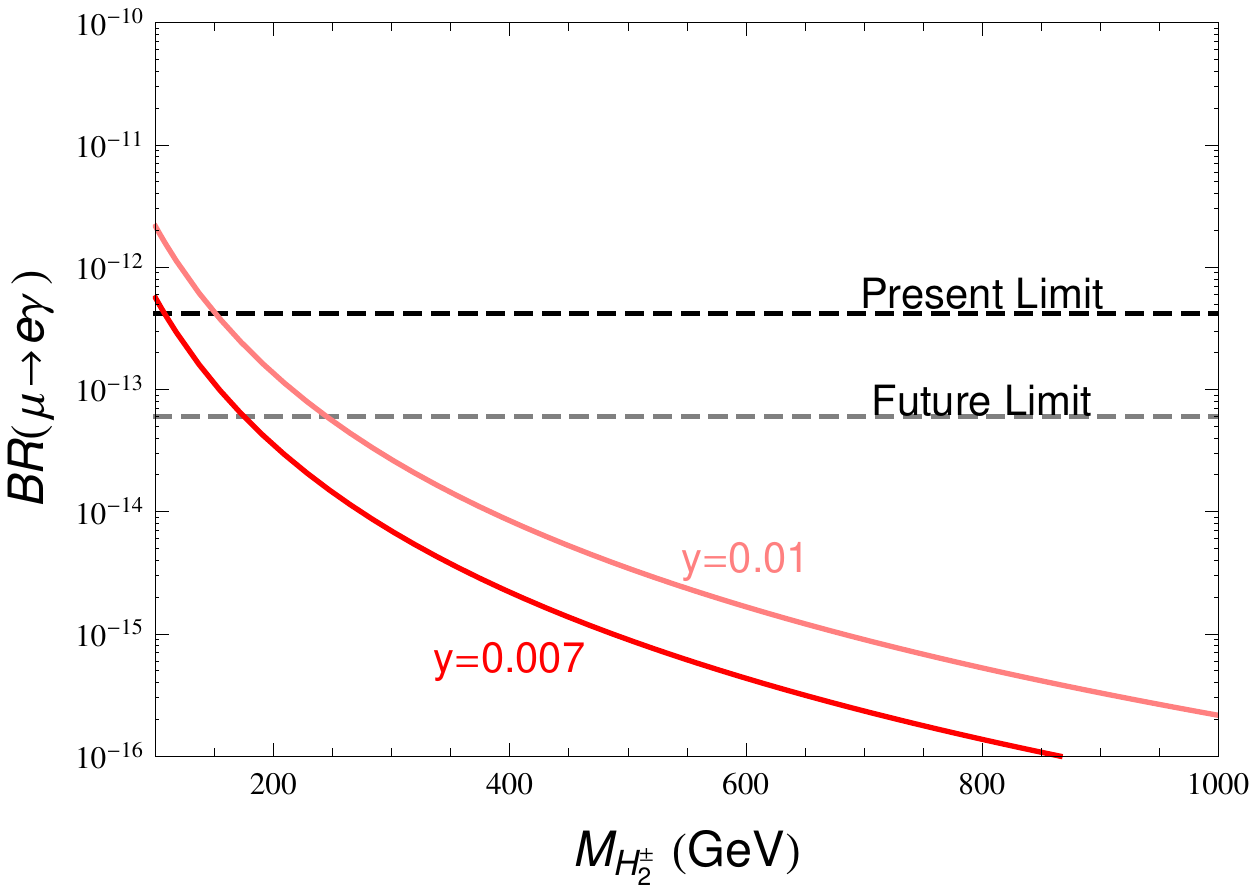}
	\caption{BR($\mu\to e\gamma$) as a function of $M_{H_2^\pm}$.}
	\label{fig:lfv}
\end{figure}
Firstly, the existence of Yukawa interaction
$\overline{L}F_{R}i\tau_{2}\eta^{\ast}$
will induce various lepton flavor violation (LFV) processes.
Detail studies on LFV processes in scotogenic models
can be found in Ref.~\cite{Toma:2013zsa}. Here, we take the current
most stringent one, i.e., the MEG experiment on the radiative decay
$\mu\to e\gamma$ with BR$(\mu\to e \gamma)<4.2\times10^{-13}$ \cite{Adam:2013mnn},
for illustration. The future limit might be down to $6\times10^{-14}$ \cite{Baldini:2013ke} .
 In the scotogenic Dirac models, the analytical
expression for branching ratio of $\mu\to e\gamma$ is calculated as
\cite{Toma:2013zsa}
\begin{equation}\label{BRLFV}
\mbox{BR}(\mu\to e\gamma)=\frac{3\alpha}{64\pi G_F^2}\left|\sum_i \frac{(y_1)_{\mu i}(y_1)_{e i}^*}{M_{H_2^+}^2} F\left(\frac{M_{F_i}^2}{M_{H_2^+}^2}\right)\right|^2,
\end{equation}
where the loop function $F(x)$ is
\begin{equation}
F(x)=\frac{1-6x+3x^2+2x^3-6x^2\ln x}{6(1-x)^4}.
\end{equation}
In Fig.~\ref{fig:lfv}, we show the BR($\mu\to e\gamma$) as a function of $M_{H_2^\pm}$ for $y=0.01,0.007$. Our BP in Eq.~\ref{BP} predicts BR($\mu\to e\gamma)\approx4\times10^{-16}$,
which is far below current and even future experimental limits.

Secondly, we briefly discuss the phenomenology of dark matter (DM).
In this paper, we mainly consider scalar DM candidate, since
for the fermion singlet, $M_F=f\langle \sigma \rangle$ is
 naturally around TeV-scale and it is
more interesting to realize successful leptogenesis.
We emphasis that the $(\Phi^{\dag}\eta)^{2}$ term is not
allowed in $U(1)_{B-L}$ extensions to generate a mass splitting
between $\eta_{R}^{0}$ and $\eta_{I}^{0}$, rendering the $\eta$
dominated component $H_2^0$ unsuitable as a DM candidate to escape the
direct detection bound. Therefore, we concentrate on the $\chi$
dominated component $H_1^0$ as the DM candidate.

\begin{figure}[!htbp]
	\centering
	\includegraphics[width=0.5\linewidth]{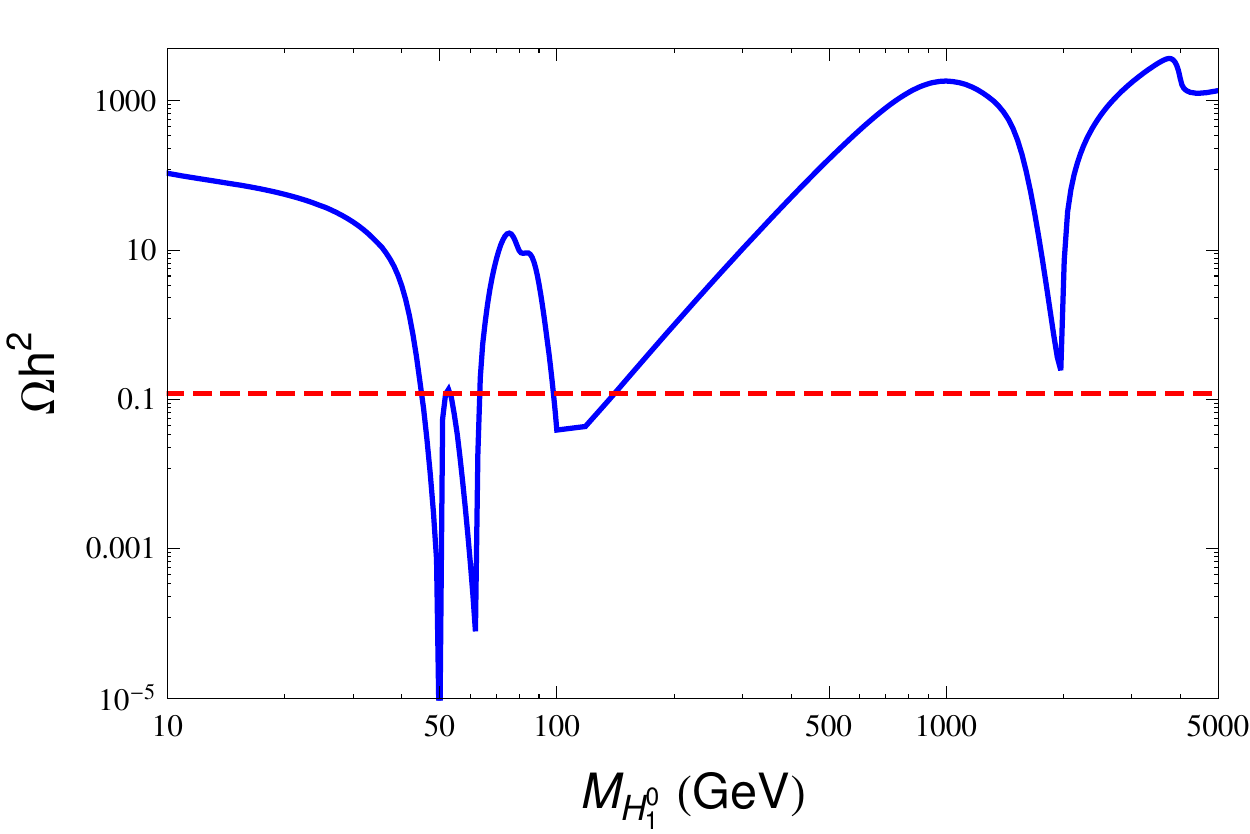}
	\caption{$\Omega h^2$ as a function of $M_{H_1^0}$. Here, we also fix $\lambda_{\Phi\chi}=\lambda_{\chi \sigma}=0.001$.}
	\label{fig:DM}
\end{figure}
With heavy $F$ and relatively small Yukawa couplings, i.e.,
$|y_2|\lesssim0.01$, the contribution of $F$ to $H_1^0$ annihilation
is negligible. To generate
the correct relic density, the possible annihilation channels are:
1) SM Higgs $h$ portal; 2) scalar singlet $H$ portal; 3) gauge boson
$Z'$ portal. For case 1), the extensive researches imply that
$M_{H_1^0}\lesssim M_h/2$ is the only allowed region under
tight constraints from relic density and direct detection
\cite{Cline:2013gha,He:2016mls}. For case 2), $M_{H_1^0}\sim M_H/2$
is needed, and electroweak scale $H_1^0$ DM is allowed
\cite{Rodejohann:2015lca}. Notably, when $M_{H}\sim100~\GeV$
thus $M_{H_1^0}\sim50~\GeV$, the observed excess in gamma-ray flux
by Fermi-LAT can be interpreted \cite{Biswas:2016ewm,DH2017}.
For case 3), it requires $M_{H_1^0}\sim M_{Z'}/2$,
and $M_{H_1^0}$ is usually around TeV-scale \cite{Klasen:2016qux}.
In Fig~\ref{fig:DM}, we show the relic density $\Omega h^2$
as a function of $M_{H_1^0}$. The Higgs $h/H$ portal could easily
acquire the correct relic density, while the $Z'$ portal could
not due to too small $g_{BL}$.  Note that the process
$H_1^0 H_1^{0*}\to HH$ could also realise correct relic density
provided $M_{H_1^0}\sim M_H$.

Thirdly, we consider Dirac leptogenesis.
It is well known that the leptogenesis can be accomplished
in Dirac neutrino models \cite{Dick:1999je,Murayama:2002je}.
In model $A_1$, the heavy Fermion singlet $F$ can decay into $L\eta$
and $\nu_R\chi$ to generate lepton asymmetry in the left-handed
$\epsilon_L$ and right-handed sector $\epsilon_R$.
Due to the fact that the sphaleron processes do not have
direct effect on right-handed fields, the lepton asymmetry in
the left-handed sector can be converted into a net baryon asymmetry
via sphaleron processes, as long as the one-loop induced
effective Dirac Yukawa couplings are small enough to prevent
the lepton asymmetry from equilibration before the
electroweak phase transition\cite{Cerdeno:2006ha}.

\begin{figure}[!htbp]
	\centering
	\includegraphics[width=0.5\linewidth]{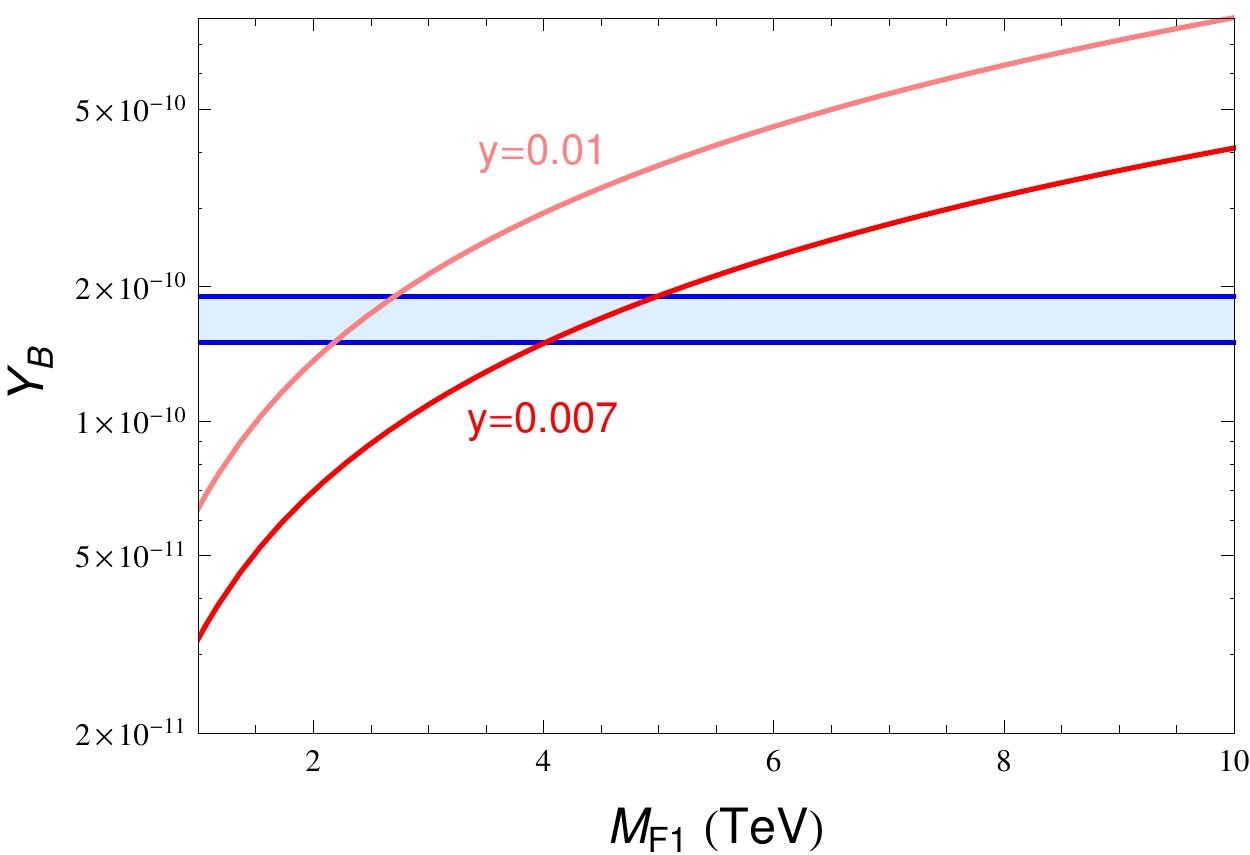}
	\caption{$Y_B$ as a function of $M_{F_1}$. The blue bound corresponds to $2\sigma$ range of Planck result. }
	\label{fig:YB}
\end{figure}

Under the assumption $y_1=y_2$, the final lepton asymmetry is calculated as \cite{Gu:2007ug}
\begin{equation}
\epsilon_{F_1}\simeq -\frac{1}{8\pi}\frac{1}{(y^\dag_1y_1)_{11}}
\sum_{j\neq1}\frac{M_{F_1}}{M_{F_j}}\text{Im}\left[(y^\dag_1
y_1)_{1j}^2\right].
\end{equation}
Define the parameter $K=\Gamma_{F_1}/H(T=M_{F_1})$, where $\Gamma_{F1}$ is the tree-level decay width of $F_1$ and $H(T)=\sqrt{8\pi^3 g_*/90}\,T^2/M_{\text{Pl}}$ wit $g_*\simeq114$ and $M_{\text{Pl}}=1.2\times10^{19}~\GeV$. As in our case $K\gtrsim1$, the final baryon asymmetry is
estimated as \cite{Cerdeno:2006ha}
\begin{equation}
Y_B=-\frac{28}{79} Y_{L_{\nu_R}}\approx- \frac{28}{79} \frac{\epsilon_{F_1}}{g_*} \frac{0.12}{K^{1.1}}.
\end{equation} 
In Fig.~\ref{fig:YB}, we depict $Y_B$ as a function of $M_{F_1}$. It is clear that the BP in Eq.~\ref{BP}
could predict the correct value of $Y_B$, as well as satisfy the out of equilibration condition
\begin{equation}
\frac{|y_1|^2|y_2|^2}{M_{F_1}}\lesssim \frac{1}{M_{\text{Pl}}} \sqrt{\frac{8\pi^3g_*}{90}}.
\end{equation} 

Then we turn to the collider phenomenology. The DM candidate $H_1^0$ will
contribute to invisible Higgs decay. The corresponding decay width for
$h\to H_1^0 H_1^{0\ast}$ is calculated as
\begin{equation}
\Gamma(h\to H_1^0 H_1^{0\ast})=\frac{g^2_{hH_1^0 H_1^{0\ast}}}{16\pi M_h}
\sqrt{1-4\frac{M_{H_1^0}^2}{M_h^2}},
\end{equation}
where $g_{hH_1^0H_1^{0\ast}}= \lambda_{\Phi \chi} v \cos\alpha + 
\lambda_{\chi\sigma} v_\sigma \sin\alpha$ 
is the effective trilinear $hH_1^0H_1^{0\ast}$
coupling and $v=246~\GeV$, $v_\sigma=M_{Z'}/(g_{BL} Q_\sigma)$. 
So the invisible branching ratio is
$\text{BR}_\text{inv}=\Gamma_\text{inv}/(\Gamma_\text{inv}+\Gamma_\text{SM})$
with $\Gamma_\text{SM}=4.07~\MeV$ at $M_h=125~\GeV$ \cite{Heinemeyer:2013tqa}.
Our BP in Eq.~\ref{BP} with $\lambda_{\Phi \chi}=\lambda_{\chi\sigma}=0.001$
 predicts $\text{BR}_\text{inv}\sim0.01$,
which can escape the most stringent bound comes from fitting to visible
Higgs decays, i.e., $\text{BR}_\text{inv}<0.23$ \cite{Khachatryan:2016vau}.
As for the light scalar $H$, the dominant visible decay is $H\to b\bar{b}$ and
invisible decay is $H\to H_1^0H_1^{0\ast}$. The possible promising
signatures are $e^+e^-\to ZH$ at future lepton colliders \cite{Baer:2013cma}.
Meanwhile, due to the doublet nature of $H_2^{\pm}$ and $H_2^0$, they can
be pair produced at LHC via Drell-Yan processes as
$pp\to H_2^+ H^-,~H_2^\pm H_2^{0(\ast)},~H_2^0H_2^{0\ast}$.
In the case of light $H_1^0$ DM, the most promising signature is
\begin{equation}
pp\to H_2^\pm H_2^{0(\ast)} \to W^\pm Z + H_1^0 H_1^{0\ast},
\end{equation}
then leptonic decays of $W$ and $Z$ will induce trilepton signature as
$2l^\pm l^- +\cancel{E}_T$. The direct searches for such trilepton
signature at LHC have excluded $M_{H_2^\pm,H_2^0}\lesssim350~\GeV$
when $M_{H_1^0}\sim50~\GeV$ \cite{Aad:2014nua,Khachatryan:2014qwa}.
\begin{figure}[!htbp]
	\centering
	\includegraphics[width=0.5\linewidth]{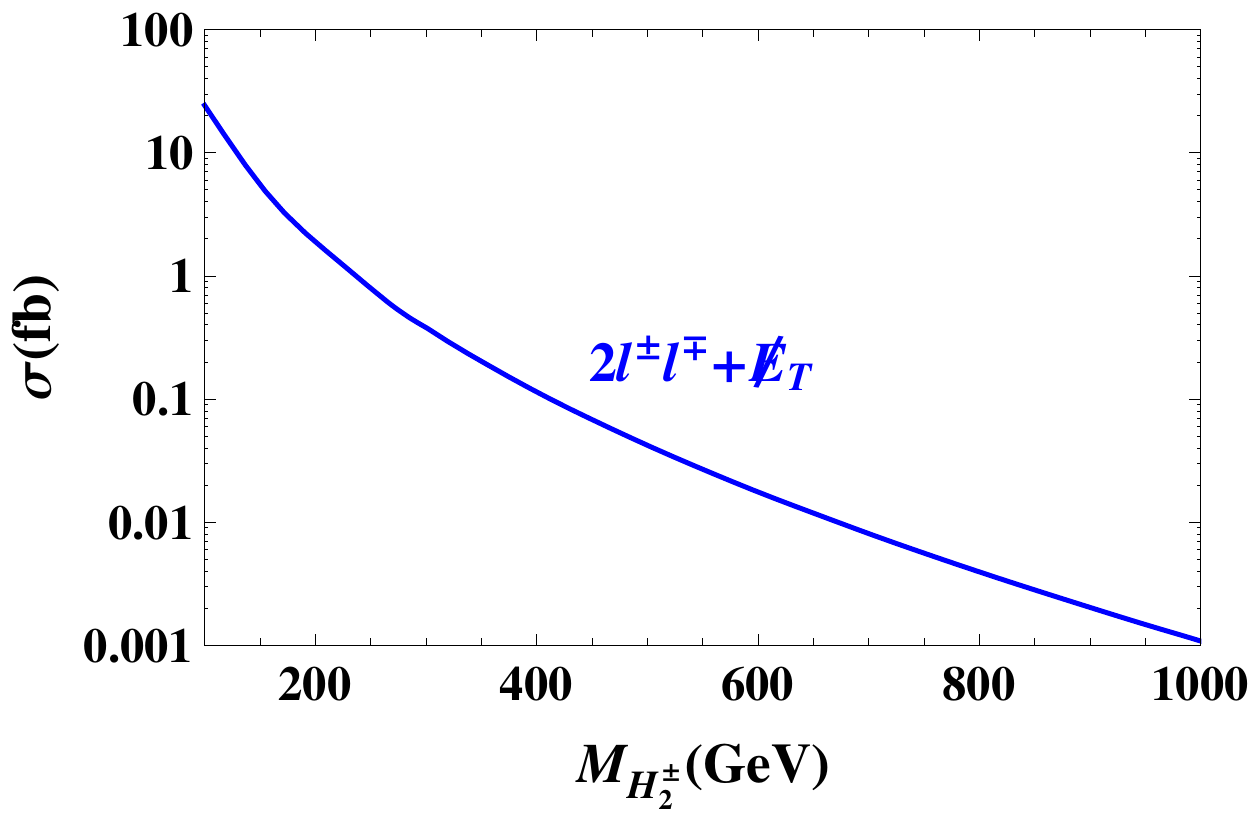}
	\caption{Trilepton signature $2\ell^\pm\ell^\mp$ as a function of $M_{H_2^{\pm}}$ at $13~\TeV$ LHC. }
	\label{fig:LHC}
\end{figure}
In Fig~\ref{fig:LHC}, we show the cross section of trilepton signature at $13~\TeV$ LHC. The cross section of our BP in Eq.~\ref{BP} is about $0.02~\fb$.

The gauged $U(1)_{B-L}$ symmetry predicts $Z^{\prime}$ boson with
mass $M_{Z^{\prime}}=Q_\sigma g_{BL}v_{\sigma}$.
Since $\sigma$ scalar is SM singlet and $
\Phi$ do not transform under $U(1)_{B-L}$, there is no mixing between $Z$ and $Z^{\prime}$ boson.
The LEP II data
requires that \cite{Cacciapaglia:2006pk}
\begin{equation}
\frac{M_{Z^\prime}}{g_{BL}}= Q_\sigma v_{\sigma}\gtrsim6\sim7~\TeV.
\end{equation}
And the direct searches for $Z'$ with SM-like gauge coupling
in the dilepton final states have excluded
$M_{Z^\prime}\lesssim4~\TeV$\cite{Aad:2014cka}. Recasting of these
searches in gauged $U(1)_{B-L}$ has been performed in
Ref.~\cite{Okada:2016gsh,Klasen:2016qux},
where the exclusion region in the
$M_{Z^\prime}-g_{BL}$ is obtained. In this paper, we consider
$M_{Z^\prime}=4~\TeV$ and $g_{BL}=0.1$ to respect these bounds.
In the limit that masses of SM fermions
$f$($f\equiv q, l, \nu_{L,R}$) are small compared with the
$Z^{\prime}$ mass, the decay width of $Z^{\prime}$ into
fermion pair $f\overline{f}$ is given by
\begin{equation}
\Gamma(Z^{\prime}\rightarrow
f\overline{f})=\frac{g_{BL}^{2}M_{Z^{\prime}}}{24\pi}C_{f}(Q_{fL}^{2}+Q_{fR}^{2})
\end{equation}
where $C_{l,\nu}=1$, $C_{q}=3$. Then the branch ratios of
$Z^{\prime}$ decay into each final states take the ratios as
\begin{equation}
\text{BR}(Z^{\prime}\rightarrow q\overline{q}):\text{BR}(Z^{\prime}\rightarrow
l^{-}l^{+}):\text{BR}(Z^{\prime}\rightarrow
\nu\overline{\nu})=4:6:3(1+Q_{\nu_{R}}^{2}),
\end{equation}
where $l=e,\mu$. Thus, the $B-L$ nature of $Z^\prime$ can be confirmed when
$\text{BR}(Z^\prime\to b\bar{b})/\text{BR}(Z^\prime\to \mu^+\mu^-)=1/3$ is
measured \cite{Kanemura:2011mw}. In addition, the decay width of $Z^{\prime}$ into
scalar pair $S S^\ast$ is given by
\begin{equation}
\Gamma(Z^\prime\to SS^\ast)=
\frac{g^2_{BL}}{48\pi} M_{Z^\prime} Q_S^2
\end{equation}
in the limit $M_S\ll M_{Z^\prime}$ as well. In case of
$H_1^0$ DM with the special mass spectrum $M_{H_1^0}<M_H
<M_{\eta^\pm,H_2^0}<M_{Z^\prime}<M_F$ as we discussed above,
the dominant invisible decays of $Z^\prime$ are $Z^\prime\to
\nu \overline{\nu}$ and $Z^\prime\to H_1^0 H_1^{0\ast}$, and
the subdominant contributions are coming from cascade decays
as $Z^\prime\to HH$ with $H\to H_1^0 H_1^{0\ast}$ and
$Z^\prime \to H_2^0 H_2^{0\ast}$ with $H_2^0\to Z(\to
\nu \overline{\nu}) H_1^0$. In Table~\ref{Zp}, we show the 
branching ratio of $Z'$ predicted by our BP.
\begin{table}[!htbp]\label{Zp}
\begin{tabular}{|c|c|c|c|c|c|c|}
\hline
$q\bar{q}$ & $\ell\bar{\ell}$ & $\nu\bar{\nu}$ & $HH$ & $H_1^0H_1^{0\ast}$ 
& $H_2^0H_2^{0\ast}$ & $H_2^+H_2^-$\\
\hline
~0.27~ & ~0.41~ & ~0.21~ & ~0.05~ & 0.02 & 0.02 & 0.02
  \\ \hline
\end{tabular}
\caption{Decay branching ratio of $Z'$.}
\end{table}
Due to different values of $B-L$
charges for the new particles in all the possible models
present in Table~\ref{ca} and Table~\ref{main1}, they
can be distinguished by precise measurement of the invisible
decays of $Z^\prime$.

\section{Conclusion}
In conclusion, we propose the $U(1)_{B-L}$ extensions of the
scotogenic models with intermediate fermion singlets added. The
Dirac nature of neutrinos is protected by $B-L$ symmetry while the
DM stability is guaranteed by the residual symmetry of $B-L$ SSB.
Under gauged $U(1)_{B-L}$, the values of $B-L$ quantum numbers for
new particles are assigned to satisfy the anomaly free condition. We
first present the topological diagrams of one-loop $Z_{2}$
realizations and subsequently check their validity under anomaly
free condition. Among the seven one-loop realizations, five of them
is available ($A_{1},A_{2},A_{4},A_{5}$ and $A_{6}$). It is found
that the total number of intermediate fermion singlets is uniquely
fixed by anomaly free condition. Especially, the $B-L$ charge
assignments for $A_{4},A_{5}$ and $A_{6}$ models can also be
uniquely fixed due to the mass splitting terms in scalar sector. We
emphasis the implications of such terms on alleviating the fine
tuning in the model and also permitting intermediate scalar singlet
as a DM candidate. Then we study the two-loop $Z_{3}$ realizations
where $n$ $F_{R/L}$ and $m$ $S_{R/L}$ fermion singlets are added.
Doing the same in one-loop model, we found $n+m$ and $B-L$ charge
assignments of all new particles is uniquely determined by anomaly
free condition. With out loss of generality, we consider the minimal
realizations with $n+m=3$ and found four viable models(denoted as
$B_{1},B_{2},B_{3}$ and $B_{4}$).

By considering phenomenology on
lepton flavor violation, dark matter, leptogenesis and LHC signatures,
we consider the benchmark point in Eq.~\ref{BP}.
In addition to generate tiny neutrino mass via scalar DM mediator,
this BP can also interpret the gamma-ray excess from
the galactic center, and realize successful leptogenesis. As for
collider signatures, the scalar DM $H_1^0$ will contribute to
invisible Higgs decay as $h\to H_1^0H_1^{0\ast}$. The scalar singlet
$H$ might be testable via $e^+e^-\to ZH$ with $H\to b\bar{b}/
H_1^0H_1^{0\ast}$ at lepton colliders. Meanwhile, the promising
signature at LHC is the trilepton signature as $pp\to
H_2^\pm H_2^{0(\ast)} \to W^\pm Z + H_1^0 H_1^{0\ast}$
with leptonic decays of $W/Z$. The new $B-L$ gauge boson is expected
discovered via the dilepton signature $pp\to Z^\prime \to
l^+l^-$ at LHC \cite{Basso:2008iv}. And in principle,
the constructed models in Table~\ref{ca} and Table~\ref{main1}
can be distinguished by precise measurement of the invisible
decays of $Z^\prime$.

\begin{acknowledgments}
The work of Weijian Wang is supported by National Natural Science
Foundation of China under Grant Numbers 11505062, Special Fund of
Theoretical Physics under Grant Numbers 11447117 and Fundamental
Research Funds for the Central Universities under Grant Numbers
 2016MS133. The work of Zhi-Long
Han is supported in part by the Grants No. NSFC-11575089.
\end{acknowledgments}

\end{document}